\documentclass[aps,pre,superscriptaddress,longbibliography]{revtex4-1}
\usepackage{multirow}
\usepackage{epsfig}
\usepackage{xcolor}
\usepackage{graphicx}% Include figure files
\usepackage{dcolumn}% Align table columns on decimal point
\usepackage{bm}% bold math
\usepackage{amsmath}

\begin{document}

\title{Data based reconstruction of complex multiplex networks}

\author{Chuang Ma}
\affiliation{School of Mathematical Science, Anhui University, Hefei 230601, China}

\author{Han-Shuang Chen}
\affiliation{School of Physics and Material Science, Anhui University, Hefei 230601, China}

\author{Xiang Li}
\affiliation{Adaptive Networks and Control Laboratory, Department
of Electronic Engineering, and the Center of Smart Networks and
Systems, School of Information Science and Engineering, Fudan University,
Shanghai 200433, China }

\author{Ying-Cheng Lai}
\affiliation{School of Electrical, Computer and Energy Engineering, Arizona State University, Tempe, Arizona 85287, USA}

\author{Hai-Feng Zhang} \email{haifengzhang1978@gmail.com}
\affiliation{School of Mathematical Science, Anhui University, Hefei 230601, China}
\affiliation{Department of Communication Engineering, North University of China, Taiyuan, Shan'xi 030051, China}

\date{\today}

\begin{abstract}
It has been recognized that many complex dynamical systems in the real world
require a description in terms of multiplex networks, where a set of common,
mutually connected nodes belong to distinct network layers and play a different
role in each layer. In spite of recent progress towards data based inference
of single-layer networks, to reconstruct complex systems with a multiplex
structure remains largely open. We articulate a mean-field based maximum
likelihood estimation framework to solve this outstanding and challenging
problem. We demonstrate the power of the reconstruction framework and
characterize its performance using binary time series from a class of
prototypical duplex network systems that host two distinct types of
spreading dynamics. In addition to validating the framework using synthetic
and real-world multiplex networks, we carry out a detailed analysis to
elucidate the impacts of structural and dynamical parameters as well as
noise on the reconstruction accuracy and robustness.

\end{abstract}
\maketitle

\section*{Introduction}

In physical and mathematical sciences, it is recognized that the
``inverse problem'' is often significantly more difficult
than the ``forward problem.'' In particular, given a system with a known
structure and a set of mathematical equations, the forward problem focuses on
analyzing and possibly solving the equations (analytically or numerically) to
uncover and understand the behaviors of the system. For the inverse problem,
the system structure and equations are unknown but only observational
or measured data are available. The task is to infer the intrinsic structure
and dynamics of the system from the data. In network science and engineering,
to reconstruct the topology of an unknown complex network and to map out
the dynamical process on the network based solely on measured time series
or data have been an active area of interdisciplinary
research~\cite{Friston:2002,GDLC:2003,PG:2003,BDLCNB:2004,
ECCBA:2005,YRK:2006,BL:2007,Timme:2007,NS:2008,DZMK:2009,PP:2009,RWLL:2010,
LP:2011,HKKN:2011,ST:2011,WLGY:2011,BHPS:2012,SBSG:2012,SWL:2012,HBPS:2013,
ZXZXC:2013,CLL:2013,TC:2014,SWL:2014,SWFDL:2014,CLL:2015,WLG:2016,JNHT:2017,
LL:2017,CL:2018,MWWHLC:2018}. A variety of approaches have been devised,
which include those based on collective
dynamics~\cite{YRK:2006,ZXZXC:2013,Timme:2007,TC:2014,NCT:2017},
stochastic analysis~\cite{MZL:2007,LL:2017}, and compressive
sensing~\cite{WYLKH:2011,SWFDL:2014,SWL:2014,MWWHLC:2018}, etc.
However, previous works focused on single-layer networks.
The goal of this paper is to address the significantly more challenging
problem of data based reconstruction of {\em multiplex} networks.

A complex system in the real world, such as the modern infrastructure, a
social or a transportation system, consists of many units connected by
different types of relationship. For example, a social network contains
different types of ties among people and a transportation system comprises
multiple types of travel platforms. Such systems require a description in terms of multiplex networks~\cite{BPPSH:2010,GBSH:2012,DSCKMPGA:2013,
KABGMP:2014,DGPA:2016}. Previous efforts in multiplex networks focused on
the forward problem to unearth the mathematical properties and the
associated physical phenomena. The main difficulty that one has to overcome
to address the inverse problem of multiplex networked systems lies in the
distinct, possibly quite diverse yet interwoven collective dynamics in
different layers. For example, the outbreak of an epidemic in the human
society induces diffusion of awareness in online social networks, leading
to two types of mutually coupled spreading dynamics~\cite{GGA:2013}, each
lives in a different network layer. Another example is that opinions can
diffuse through different channels (layers) and interact with each other.
To our knowledge, there has been no prior work on the inverse problem
for multiplex networks.

In this paper, we develop a reconstruction framework based on mean-field
maximum likelihood estimation to address the problem of data based
reconstruction of multiplex networks. To describe our framework in a
concrete and clear way, we focus on duplex networked systems - perhaps the
most extensively studied multiplex networks that are relevant to real world
situations such as complex cyberphysical systems. We assume that each layer
hosts a distinct type of spreading dynamics and the two types of processes
are interwoven. In particular, one (physical) layer hosts the
susceptible-infected-susceptible (SIS) type of spreading dynamics, while
the other (virtual) layer is a social network with information spreading
governed by the unaware-aware-unaware (UAU) process~\cite{GGA:2013}.
Provided that binary time series data are available from both layers, we
show that our framework is capable of accurately reconstructing the full
topology of each layer for a large number of empirical and synthetic networks.
We elucidate the impacts of network structural and dynamics parameters
on the reconstruction accuracy, such as the average degree, interlayer
coupling, and heterogeneity in the spreading rates. The effect of noise is
also investigated. Our framework represents a success to assess the ``internal
gear'' of complex systems with a duplex structure, which can be generalized
to networks with a more sophisticated structure and diverse dynamical
processes.

\section*{Results}

\paragraph*{Basic notions of reconstruction framework.}
Duplex networks with UAU-SIS coupled dynamics describe realistic situations
where there is competition between disease spreading and social awareness,
with the physical contact layer supporting an epidemic process and the virtual
contact layer generating awareness diffusion, where the former is of the
SIS type while the latter can be described by the UAU process.

The mutual interplay or coupling between the two distinct types of spreading
processes can be described, as follows. An I-node in the physical layer is
automatically aware of the infection and changes to the A (aware) state
immediately in the virtual layer. If an S-node is in the A state, its ability
to infect other connected nodes is discounted by a factor $0 \leq \gamma < 1$.
For convenience, we introduce the mathematical notation
${}_{{x_4}}^{{x_3}}X_{{x_1}}^{{x_2}}$ associated with a variable $X$, where
$x_1$ and $x_2$ are the time attribute and the id number of $X$ (e.g.,
$x_1=t$ and $x_2=i$), $x_3$ determines whether $X$ belongs to the virtual
layer (i.e., $x_3=1$) or the physical contact layer (i.e., $x_3=2$), $x_4$
specifies either U or A. For example, $\sideset{_U}{}{\mathop{\beta}}$ and
$\sideset{_A}{}{\mathop{\beta}}=\gamma\cdot\sideset{_U}{}{\mathop{\beta}}$
denote the infection rate of an unaware and an aware S-node, respectively.
Variable $s$ with three annotations denotes the state of a node, e.g.,
${}^{{1}}s_{{t_m}}^{{i}}=0$ or 1 (${}^{{2}}s_{{t_m}}^{{i}}=0$ or 1) indicates
that node $i$ in the virtual (physical) layer is in U or A state (S or I state)
at time $t_m$. The framework can deal with the general case where the
transmission and/or recovery rates are heterogeneous among the nodes.

Our reconstruction framework consists of three steps: (a) to formulate the
likelihood function of the coupled dynamics, (b) to perform the mean-field
based approximation to enable the estimation, and (c) to transform the
estimation problem into two solvable linear systems - one for each layer
with solutions representing the neighbors of each node in the layer.
A detailed formulation of the reconstruction framework is presented in
{\bf Methods} and a schematic illustration of the UAU-SIS coupling dynamics
is given in Fig.~\ref{figS:sketch} of Supplementary Information (SI).

\begin{figure}
\centering
\includegraphics[width=\linewidth]{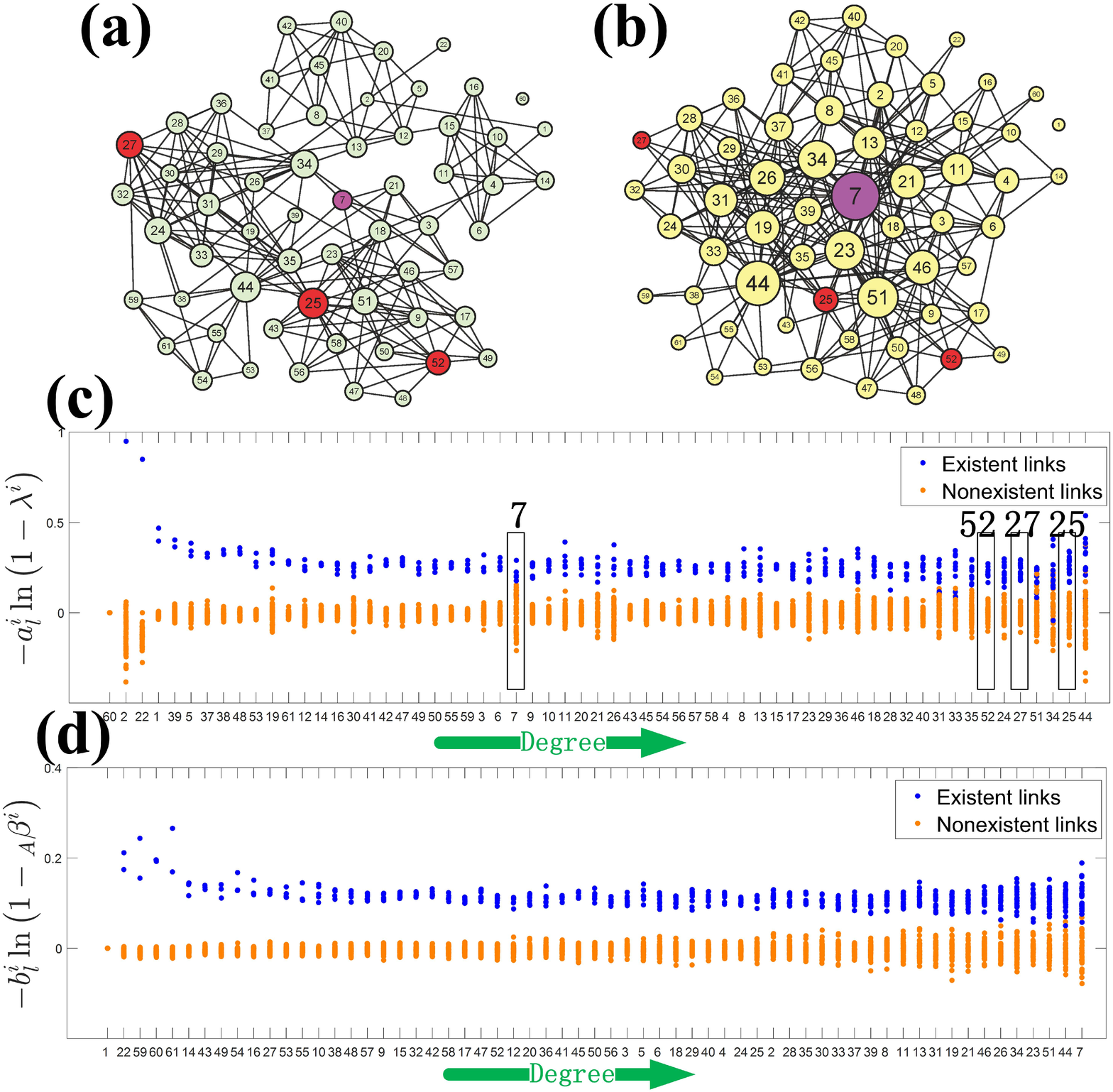}
\caption{ {\bf Reconstruction of the CS-AARHUS network}. (a) Actual structure
of the virtual contact layer (Facebook). (b) The structure of the physical
layer. (c,d) The values of $-a_l^i\ln\left({1-{\lambda^i}}\right),~i\neq l$ and
$-b_l^i\ln \left({1-{\lambda^i}}\right)~i\neq l$, respectively, versus the
nodal degree. Each column gives the connectivity of a node, where the blue
and orange dots denote the existent and nonexistent links, respectively.
The length of the time series is $M=30000$. The parameter values for
the dynamical processes are $\lambda=0.2$, ${}_U{\beta}=0.2$,
${}_A{\beta}=0.5{}_U{\beta}$, and $\mu=\delta=0.8$.}
\label{fig:double-layer}
\end{figure}

\begin{figure}
\centering
\includegraphics[width=\linewidth]{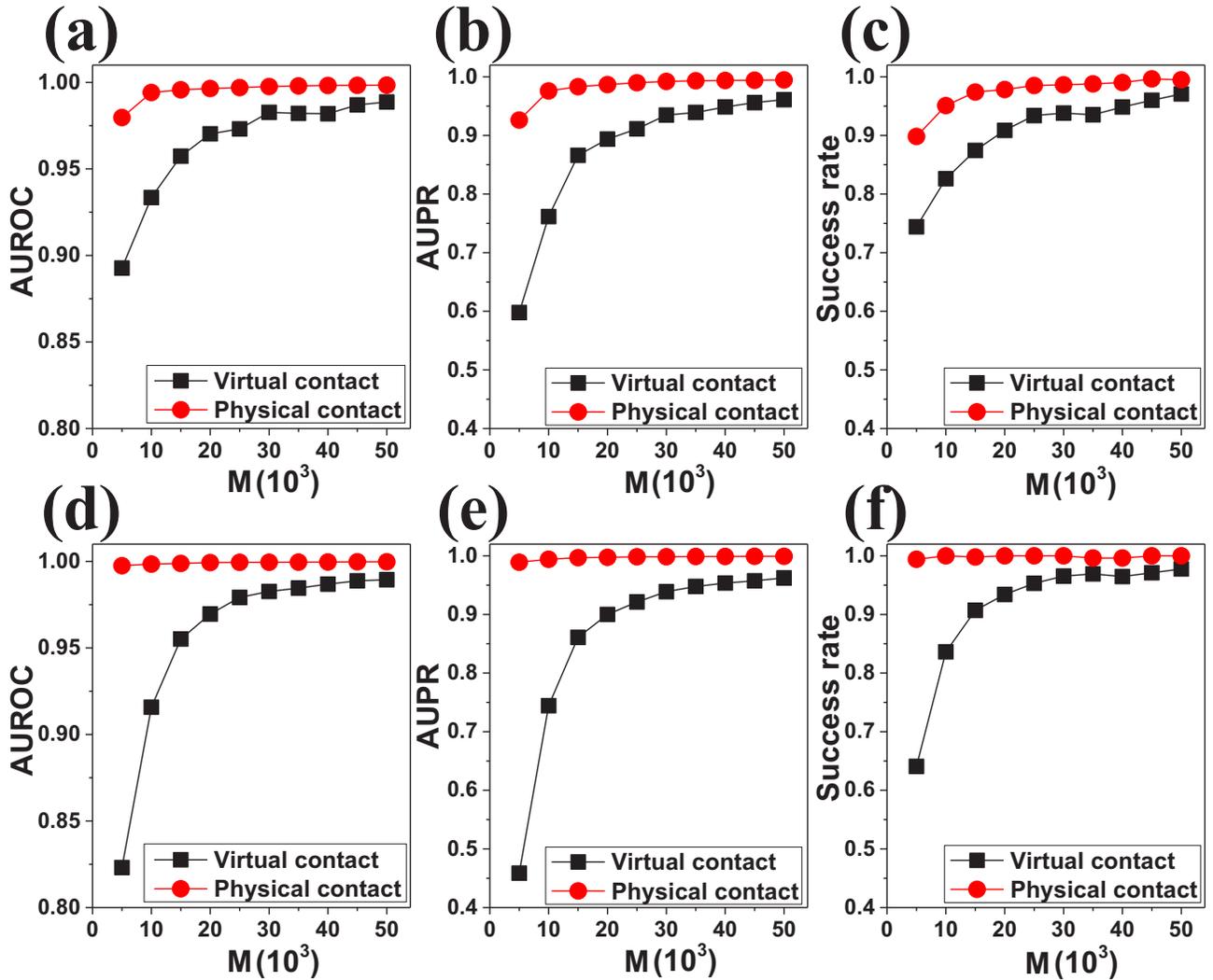}
\caption{{\bf Reconstruction accuracy of C.~elegans and CKM networks}.
(a-c) For the C.~elegans network, values of AUROC, AUPR and Success rate
versus the length $M$ of the time series, respectively, for the parameter
setting $\lambda=0.2$, ${}_U{\beta}=0.3$, ${}_A{\beta}=0.5{}_U{\beta}$,
and $\mu=\delta=0.8$. (d-f) The corresponding results for the CKM network
for the same parameter values as for (a-c) except ${}_U{\beta}=0.5$.}
\label{fig:two_empirical}
\end{figure}

\begin{figure}
\centering
\includegraphics[width=0.8\linewidth]{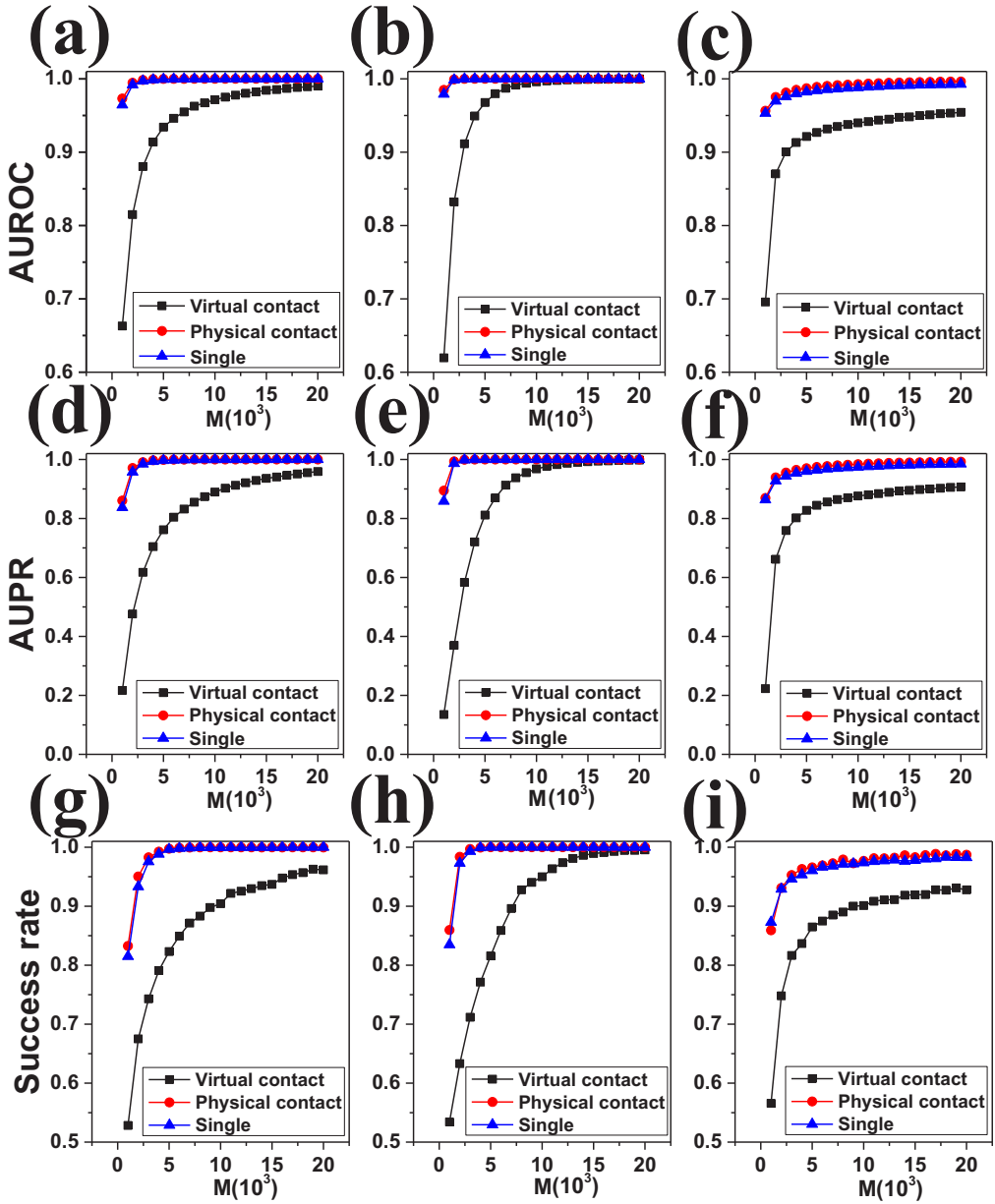}
\caption{ {\bf Effect of interlayer coupling on reconstruction accuracy}.
Columns 1-3: reconstruction performance for ER-ER, SW-SW, and BA-BA duplex
networks, respectively. The ``single'' case indicates the absence of
interlayer coupling: ${}_U{\beta}=0$ ($\lambda=0$) for the virtual (physical)
layer. The parameter setting is $\lambda=0.4$, ${}_U{\beta}=0.4$,
${}_A{\beta}=0.5{}_U{\beta}$, and $\mu=\delta=0.8$. The structures of the
two layers are identical. The network parameters are $N=100$ and
$\langle k_1\rangle=\langle k_2\rangle=6$.}
\label{fig:interact}
\end{figure}

\begin{figure}
\centering
\includegraphics[width=\linewidth]{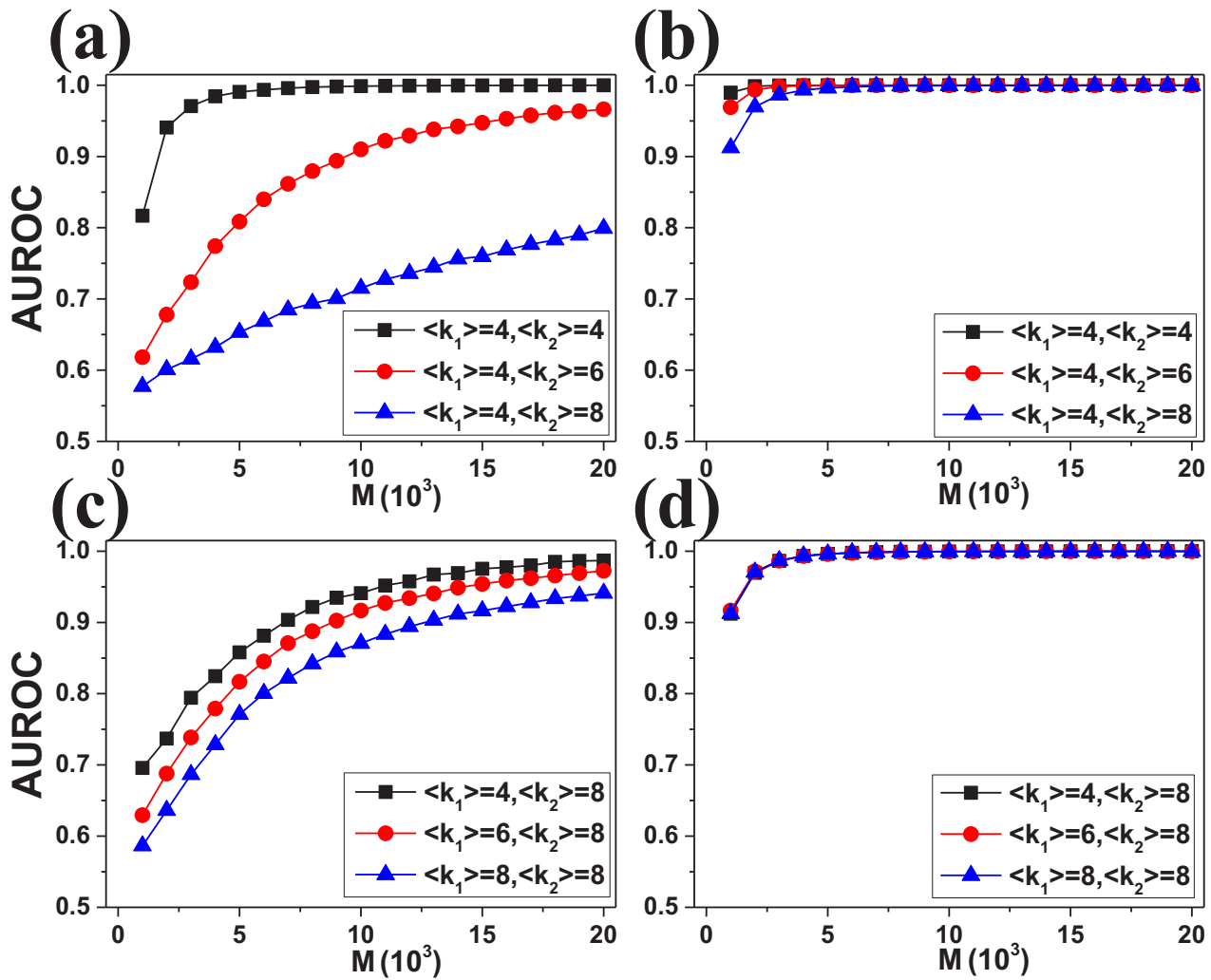}
\caption{ {\bf Effect of average degree on reconstruction as measured by the
AUROC index}. (a,b) For a fixed value of the average degree
$\langle k_1\rangle$ of the virtual layer, the effect of varying the average
degree $\langle k_2\rangle$ of the physical layer on the reconstruction
accuracy of the former and latter, respectively. (c,d) For a fixed value
of $\langle k_2\rangle$, the effect of varying the value of
$\langle k_1\rangle$ on the reconstruction accuracy of the virtual and physical
layer, respectively. ER-ER duplex networks with $N$=100 are used. The
parameters are $\lambda=0.3$, ${}_U{\beta}=0.4$, ${}_A{\beta}=0.5{}_U{\beta}$,
and $\mu=\delta=0.8$.}
\label{fig:ave_degree}
\end{figure}

\begin{figure}
\centering
\includegraphics[width=\linewidth]{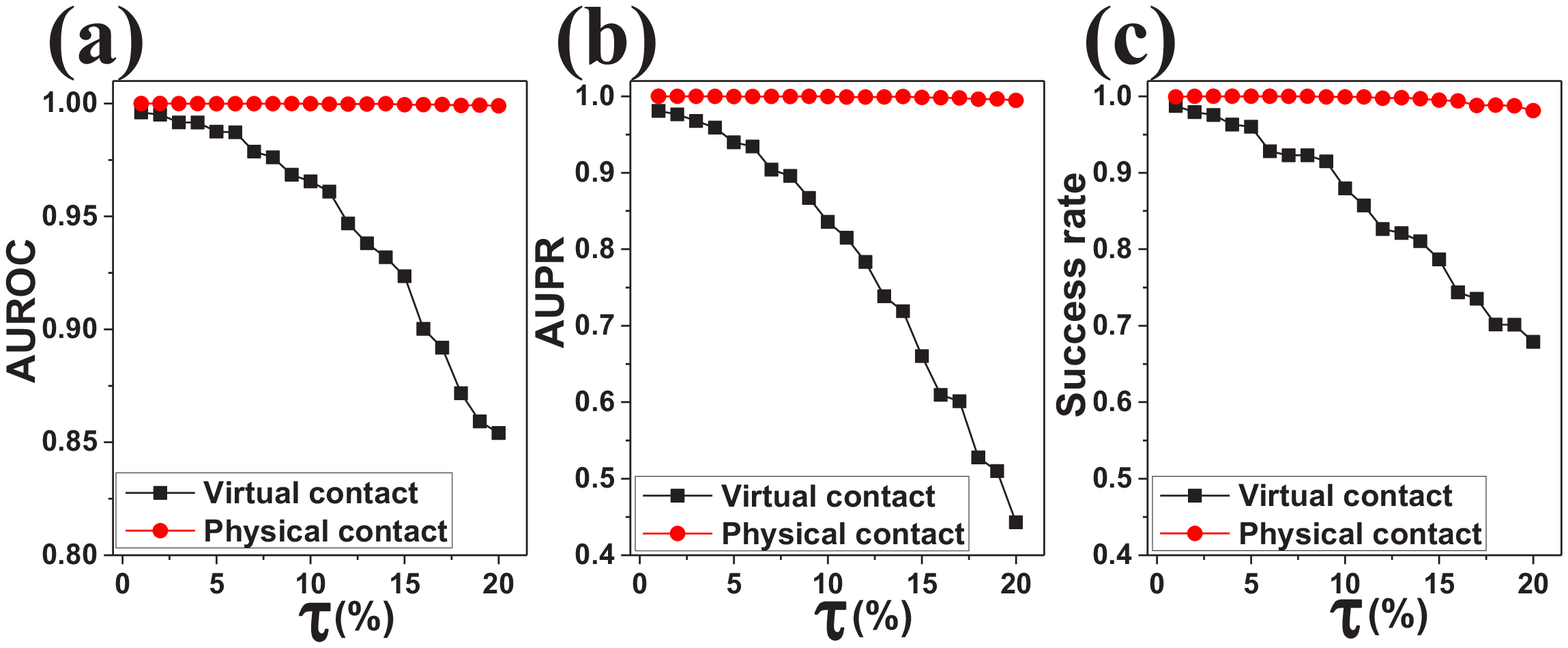}
\caption{ {\bf Impact of noise on reconstruction accuracy}. (a-c) AUROC, AUPR
and Success rate versus the fraction $\tau$ of randomly flipped states for an
ER-ER duplex system. The network parameters are $N=100$,
$\langle k_1\rangle=4$, and $\langle k_2\rangle=6$. The length of the time
series is $M=30000$. Other parameters are the same as in
Fig.~\ref{fig:ave_degree}.}
\label{fig:noise}
\end{figure}

\paragraph*{Performance demonstration: reconstruction of three real-world
duplex networks.}
For a network layer (say the virtual layer), the mathematical formulation of
our reconstruction framework gives a characteristic, node-specific quantity
for distinguishing the existent from the non-existent links:
$-a_l^i\ln \left({1-{\lambda^i}}\right)$, where
$\mathbf{a}^i$ is the vector defining the local connection structure of
node $i$: $a^i_l=1$ specifies that node $l$ is in the neighborhood of $i$
and $a^i_j=0$ otherwise, and $\lambda^i$ is the probability that node
$i$ (if it is unaware) is informed by an aware (A) neighbor. For actual
links, the values of $-a_l^i\ln \left({1-{\lambda^i}}\right)$ are finite
and away from zero, while the values for non-existent links are near
zero. If there is a gap between the two sets of values, the actual links
can be detected. A similar quantity can be defined for the physical layer,
which is denoted as $-b_l^i\ln \left(1 -{}_A\beta \right)$.

We first validate our framework using an empirical network of 61 nodes, the
so-called CS-AARHUS network~\cite{MMR:2013}. The original network has five
layers. We regard the Facebook layer as the virtual layer and the other four
off-line layers (Leisure, Work, Co-authorship, Lunch) as the physical layer,
as illustrated in Figs.~\ref{fig:double-layer}(a,b), respectively.
Figures~\ref{fig:double-layer}(c,d) show the values of characteristic
quantities $-a_l^i\ln \left({1-{\lambda^i}}\right)$ and
$-b_l^i\ln \left(1 -{}_A\beta \right)$ for the virtual and physical layers,
where the blue and orange dots denote the actual and nonexistent links,
respectively. We see that the values of the characteristic quantities
are well separated by a distinct gap and can be unequivocally distinguished
through a properly chosen threshold (see also Fig.~\ref{fig:cutoff} in SI). For the physical
layer in Fig.~\ref{fig:double-layer}(d), the gap between the blue and orange
dots exhibits a decreasing trend with the nodal degree, indicating that the
neighbors of larger degree nodes are harder to be detected because of
neighborhood overlapping associated with such nodes. This result is consistent
with previous findings~\cite{SWFDL:2014,MZL:2007}. For the virtual layer
[Fig.~\ref{fig:double-layer}(c)], the blue and orange dots for node 7 are
overlapped even though ${}^1{k^{7}}=6$, but there is a finite gap for large
degree nodes, e.g., node 52 with ${}^1{k^{52}}=10$, node 27 with
${}^1{k^{27}}=12$, and node 25 with ${}^1{k^{25}}=15$. The relatively small
gap of ${}^1{k^{7}}$ is due to the fact that the counterpart value in the
physical layer is large: ${}^2{k^{7}}=29$, indicating that the node has been
infected and is thus constantly in the A state in the virtual layer (an
infected node becomes aware immediately). As a result, the states of the
neighbors of this node in the virtual layer have little influence on its
state, making reconstruction difficult. For nodes with large and small
degrees in the virtual and physical layers, respectively, the transition
from U to A is mainly determined by the states of the neighbors, facilitating
reconstruction. In general, the structure of the physical layer has a
significant effect on the reconstruction of the virtual layer, but the
effect in the opposite direction is minimal.

We next demonstrate the power of our reconstruction framework for two
additional empirical networks: C.~elegans~\cite{DGPA:2016} and the innovation
diffusion network for physicians (the so-called CKM network)~\cite{CKM:1957}.
Each network has three layers. We choose the first and the second layer
as the virtual and physical layer, respectively. The panels in the top and
bottom rows of Fig.~\ref{fig:two_empirical} display the reconstruction
accuracy in terms of the statistical quantities AUROC, AUPR and Success rate
({\bf Methods}) versus the length of the time series for the C.~elegans and
CKM networks, respectively. We have that longer time series result in better
reconstruction performance and the reconstruction accuracy of the physical
layer is higher than that of the virtual layer, which are consistent
with the results in Fig.~\ref{fig:double-layer}.

\paragraph*{Performance analysis: reconstruction of synthetic multiplex
networks.}
To understand the effect of interlayer coupling on reconstruction,
we test a number of synthetic complex networks: small-world
(SW-SW)~\cite{WS:1998}, Erd\"{o}s-R\'{e}nyi (ER-ER)~\cite{ER:1960}, and
Barab\'asi-Albert (BA-BA)~\cite{BA:1999} duplex systems. For comparison, we
include the special case where each layer is separately reconstructed without
taking into account the other layer, which is equivalent to reconstructing a
single-layer network (labeled as single). Figures~\ref{fig:interact}(a-i) show
that the reconstruction accuracy of the virtual layer is greatly reduced when
a physical layer is introduced (e.g., blue/red $\rightarrow$ black).
Without the physical layer, the transition of an unaware node in the virtual
layer to the aware state depends only on the states of its neighbors. With the
presence of the physical layer, an A-node can spontaneously become aware
once it is infected, ``concealing'' the information about the structure of
the virtual layer. On the contrary, the reconstruction accuracy of the
physical layer can be improved slightly (e.g., blue versus red) when
the virtual layer is introduced, which reduces the ability to infect of A-nodes and
prevents too many nodes from being in the I state, facilitating reconstruction.
Figure ~\ref{fig:interact} also illustrates that the reconstruction accuracy
of the SW-SW duplex network is higher than that of the ER-ER duplex network
and much higher than that of the BA-BA duplex network due to the difficulty
in reconstructing the neighbors of large degree nodes.

How does the average degree of each layer affect the reconstruction accuracy?
Figure~\ref{fig:ave_degree}(a) shows that an increase in the average degree
$\langle k_2\rangle$ of the physical layer can greatly reduce the
reconstruction accuracy of the virtual layer. Figures~\ref{fig:ave_degree}(b,c)
show that, for the physical layer, the accuracy gradually decreases with its
average degree, for a fixed average degree of the virtual layer. An explanation
is that the probability of being infected tends to increase for a larger value
of $\langle k_1\rangle$, ``hiding'' the information required for uncovering
the structure of the virtual layer. We also find that increasing the average
degree $\langle k_1\rangle$ of the virtual layer tends to reduce the
reconstruction accuracy of itself [Fig.~\ref{fig:ave_degree}(c)] but has a
negligible effect on the reconstruction of the physical layer
[Fig.~\ref{fig:ave_degree}(d)].

Figure~\ref{fig:noise} shows the effect of noise on the reconstruction
accuracy, where noise is implemented by randomly flipping a fraction $\tau$
of the states among the total number $MN$ of states. Noise has a significant
effect on the reconstruction of the virtual layer, but it hardly affects the
reconstruction of the physical layer (even when the flip rate is
$\tau = 20\%$).

\section*{Discussion}

We have developed a mean-field based maximum likelihood estimation framework
to solve the challenging problem of data based reconstruction of multiplex
networks. The reconstruction performance has been demonstrated using a number
of real-world and synthetic duplex networks comprising a virtual and a
physical layer, where each layer hosts a distinct type of spreading dynamics
that are coupled through the duplex network structure. Extensive test and
analysis indicate that the framework is capable of accurately reconstructing
the full topology of each layer based solely on measured time series. A
thorough examination of the dynamical coupling between the two layers gives
that the reconstruction accuracy of the physical layer is generally much
higher than that of the virtual layer. In addition, the reconstruction
accuracy of the virtual layer is more sensitive to external noise than
that of the physical layer.

Our framework represents a starting point towards reconstructing more general
multiplex networks hosting different types of dynamics. Appealing features
are that the framework is parameter free, has high accuracy, is readily to
be implemented, and has a solid mathematical foundation. Issues warranting
further considerations include extension to continuous-time dynamical
processes, generalization to multiplex networks consisting of more than
two layers, and development of effective and practical methods to reduce
the required data amount.

\section*{Methods}

\paragraph*{ UAU-SIS dynamics on duplex networks.}
The UAU-SIS model was originally articulated to study the competition between
social awareness and disease spreading on double layer networks, where the
physical contact layer supports an epidemic process and the virtual contact
layer supports awareness diffusion~\cite{GGA:2013}. The two layers share
exactly the same set of nodes but their connection patterns are different.
Spreading of awareness in the virtual layer is described by the classic SIS
type of epidemic dynamics: an unaware node (U) is informed by an aware
neighbor (A) with probability $\lambda$, and an aware node can lose awareness
and returns to the U state with probability $\delta$. Epidemic dynamics in the
physical layer are also of the SIS type, where an infected (I) node can infect
its susceptible (S) neighbors with probability $\beta$, and an I-node returns
to the S state with probability $\mu$. The interlayer interaction is
described, as follows. An I-node in the physical layer is automatically aware
of the infection and changes to the A state immediately in the virtual layer.
If an S-node is in the A state, its ability to infect is reduced by a factor
$0\leq\gamma<1$. Figure~\ref{figS:sketch} in SI presents a schematic illustration of the
duplex network with the described interacting dynamical process.

\paragraph*{Mathematical formulation of the reconstruction framework.}
For a duplex network hosting UAU-SIS dynamics, at any time a node can be in
one of the three distinct states: unaware and susceptible (US), aware and
susceptible (AS), or aware and infected (AI). The state UI (unaware and
infected) is redundant because an I-node becomes aware immediately. The
connections of node $i$ in the virtual and physical layers are specified by
the vectors $\mathbf{a}^i$ and $\mathbf{b}^i$, respectively, where $a^i_j=1$
indicates that node $j$ is a neighbor of node $i$ in the virtual layer and
$a^i_j=0$ otherwise, and $b^i_j$ is defined similarly.

%The numbers of
%A-neighbors in the virtual layer and I-neighbors in the physical layer are
%$\sum\limits_{j\ne i}{a_j^i~{}^1s_t^j}$ and
%$\sum\limits_{j\ne i}{b_j^i~{}^2s_t^j}$, respectively. In the absence of
%dynamical correlation, the relevant probabilities are
%\begin{eqnarray}
%\nonumber
%r_t^i & = & {\left({1-{\lambda^i}}\right)^{\sum\limits_{j\ne i}{a_j^i~{}^1s_t^j}}}, \\ \nonumber
%{}_Uq_t^i & = & {\left({1-{}_U{\beta^i}}\right)^{\sum\limits_{j\ne i}{b_j^i~{}^2s_t^j} }}, \\ \nonumber
%{}_Aq_t^i & = & {\left({1-{}_A{\beta^i}}\right)^{\sum\limits_{j\ne i}{b_j^i~{}^2s_t^j} }},
%\end{eqnarray}
%where $r_t^i$ is the probability that node $i$ is not informed by any
%neighbor, ${}_Uq_t^i$ and ${}_Aq_t^i$ are the probabilities that node $i$
%is not infected by any neighbor if it was in the U state and A state,
%respectively.

The various transition probabilities can be derived from the transition
tree of the UAU-SIS model (Fig.~\ref{fig:trans-tree} in SI). Let
$\sideset{_U}{}{\mathop{\beta}}$ and
$\sideset{_A}{}{\mathop{\beta}}=\gamma\cdot\sideset{_U}{}{\mathop{\beta}}$
be the infection rates of unaware and aware S-node, respectively.
For node $i$, $\sum\limits_{j \ne i} {a_j^i~{}^1s_t^j}$ and
$\sum\limits_{j \ne i} {b_j^i~{}^2s_t^j}$ are the numbers of A-neighbors
in the virtual layer and I-neighbors in the physical layer, respectively.
The following probabilities are essential to the spreading dynamics:
(1) $r_i^t$, the probability that node $i$ is not informed by any neighbor,
(2) ${}_Uq_t^i$, the probability that node $i$ is not infected by any
neighbor if $i$ was unaware, and (3) ${}_Aq_t^i$, the probability that
node $i$ is not infected by any neighbor if $i$ was aware. In the absence
of any dynamical correlation, these probabilities are given by
\begin{eqnarray} \label{eq:transition_main}
\begin{array}{l}
r_t^i = {\left( {1 - {\lambda ^i}} \right)^{\sum\limits_{j \ne i} {a_j^i{}^1s_t^j} }},\\
{}_Uq_t^i = {\left( {1 - {}_U{\beta ^i}} \right)^{\sum\limits_{j \ne i} {b_j^i{}^2s_t^j} }},\\
{}_Aq_t^i = {\left( {1 - {}_A{\beta ^i}} \right)^{\sum\limits_{j \ne i} {b_j^i{}^2s_t^j} }}.
\end{array}
\end{eqnarray}
A tacit assumption in Ref.~\cite{GGA:2013} is that diffusion of awareness in
the virtual layer occurs before epidemic spreading in the physical layer. In
our work, we do not require that the two types of spreading dynamics occur in
any particular order.

From Fig.~\ref{fig:trans-tree} in SI and Eq.~(\ref{eq:transition_main}), we have that the
probabilities of node $i$ being in the US, AS and AI states at $t+1$ when it
is in the US state at time $t$ are
\begin{eqnarray} \label{eq:US_main}
\begin{array}{l}
P^{US\rightarrow US} = r_t^i{}_Uq_t^i,\\
P^{US\rightarrow AS} = \left( {1 - r_t^i} \right){}_Uq_t^i,\\
P^{US\rightarrow AI} = r_t^i\left( {1 - {}_Uq_t^i} \right) + \left( {1 - r_t^i} \right)\left( {1 - {}_Uq_t^i} \right) = 1 - {}_Uq_t^i.
\end{array}
\end{eqnarray}
The probabilities of node $i$ being in the US, AS and AI states at $t+1$ when
it is in the AS state at time $t$ are
\begin{eqnarray} \label{eq:US_1_main}
\begin{array}{l}
P^{AS\rightarrow US} = {\delta ^i}{}_Aq_t^i, \\
P^{AS\rightarrow AS}= \left( {1 - {\delta ^i}} \right){}_Aq_t^i, \\
P^{AS\rightarrow AI} = {\delta ^i}\left( {1 - {}_Aq_t^i} \right) +
\left( {1 - {\delta ^i}} \right)\left( {1 - {}_Aq_t^i} \right) = 1 - {}_Aq_t^i.
\end{array}
\end{eqnarray}
The probabilities for node $i$ to be in the US, AS and AI states at $t+1$
if it is in the AI state at time $t$ are
\begin{eqnarray} \label{eq:AI_main}
\begin{array}{l}
P^{AI\rightarrow US} = {\delta ^i}{\mu ^i}, \\
P^{AI\rightarrow AS}= \left( {1 - {\delta ^i}} \right){\mu ^i}, \\
P^{AI\rightarrow AI} = 1 - {\mu ^i}.
\end{array}
\end{eqnarray}
Say only the states ${}^1s_{t_{m}}^i$ and ${}^2s_{t_m}^i$ ($i=1,\ldots,N$)
at time $t_m$ (not necessarily uniform) are recorded, where $N$ is the network
size. Our framework consists of three steps: (1) to reconstruct the
likelihood function of the coupled dynamics, (2) to apply the mean-field
approximation to enable maximum likelihood estimation (MLE), and (3) to
transform the MLE problem into two solvable linear systems - one for each
layer with solutions representing the neighbors of each node in the layer.
In particular, the likelihood function of node $i$ is given by (see
Sec.~III of SI):
\begin{widetext}
\begin{eqnarray} \label{eq:likelihood_main}
\begin{array}{l}
\quad P\left( {{{\left\{ {{}^1s_{{t_m} + 1}^i,{}^2s_{{t_m} + 1}^i} \right\}}_{m = 1 \cdots M}}|{{\left\{ {{}^1s_{{t_m}}^j,{}^2s_{{t_m}}^j} \right\}}_{j = 1 \cdots N,m = 1 \cdots M}},{\mathbf{a}^i},{\mathbf{b}^i},{\lambda ^i},{}_U{\beta ^i},{}_A{\beta ^i},{\delta ^i},{\mu ^i}} \right)\\
 = {\prod\limits_m {\left[ {{{\left( {r_{{t_m}}^i{}_Uq_{{t_m}}^i} \right)}^{\left( {1 - {}^1s_{{t_m} + 1}^i} \right)\left( {1 - {}^2s_{{t_m} + 1}^i} \right)}}{{\left( {\left( {1 - r_{{t_m}}^i} \right){}_Uq_{{t_m}}^i} \right)}^{{}^1s_{{t_m} + 1}^i\left( {1 - {}^2s_{{t_m} + 1}^i} \right)}}{{\left( {1 - {}_Uq_{{t_m}}^i} \right)}^{{}^1s_{{t_m} + 1}^i{}^2s_{{t_m} + 1}^i}}} \right]} ^{\left( {1 - {}^1s_{{t_m}}^i} \right)\left( {1 - {}^2s_{{t_m}}^i} \right)}}\\
\quad \quad {\left[ {{{\left( {{\delta ^i}{}_Aq_{{t_m}}^i} \right)}^{\left( {1 - {}^1s_{{t_m} + 1}^i} \right)\left( {1 - {}^2s_{{t_m} + 1}^i} \right)}}{{\left( {\left( {1 - {\delta ^i}} \right){}_Aq_{{t_m}}^i} \right)}^{{}^1s_{{t_m} + 1}^i\left( {1 - {}^2s_{{t_m} + 1}^i} \right)}}{{\left( {1 - {}_Aq_{{t_m}}^i} \right)}^{{}^1s_{{t_m} + 1}^i{}^2s_{{t_m} + 1}^i}}} \right]^{{}^1s_{{t_m}}^i\left( {1 - {}^2s_{{t_m}}^i} \right)}}\\
\quad \quad {\left[ {{{\left( {{\delta ^i}{\mu ^i}} \right)}^{\left( {1 - {}^1s_{{t_m} + 1}^i} \right)\left( {1 - {}^2s_{{t_m} + 1}^i} \right)}}{{\left( {\left( {1 - {\delta ^i}} \right){\mu ^i}} \right)}^{{}^1s_{{t_m} + 1}^i\left( {1 - {}^2s_{{t_m} + 1}^i} \right)}}{{\left( {1 - {\mu ^i}} \right)}^{{}^1s_{{t_m} + 1}^i{}^2s_{{t_m} + 1}^i}}} \right]^{{}^1s_{{t_m}}^i{}^2s_{{t_m}}^i}},
\end{array}
\end{eqnarray}
\end{widetext}
where $M$ is the length of the time series (step 1). The logarithmic form of
Eq.~(\ref{eq:likelihood_main}), denoted as
\begin{displaymath}
L = L\left( {{a^i},{b^i},{\lambda ^i},{}_U{\beta ^i},{}_A{\beta ^i},
{\delta ^i},{\mu ^i}} \right),
\end{displaymath}
can be written as
\begin{displaymath}
L = {L_0}\left( {{\delta ^i},{\mu ^i}} \right) + {L_1}
\left({{\mathbf{a^i}},{\lambda^i}}\right)+{L_2}
\left({{\mathbf{b^i}},{}_U{\beta ^i},{}_A{\beta ^i}}\right).
\end{displaymath}
MLE can be performed by maximizing $L_0$, $L_1$ and $L_2$ (Sec.~III of SI).
Because $L_0$ does not contain information about the network structure, it
is only necessary to maximize $L_1$ and $L_2$ with respect to $a^i_j$ and
$b^i_j$, respectively. The likelihood function for reconstructing the
virtual layer is (see Sec.~III of SI for the corresponding function for
the physical layer):
\begin{eqnarray} \label{eq:virtual_like_main}
{L_1}\left( {{\mathbf{a^i}},{\lambda ^i}} \right)  =
\sum\limits_m {}^1X_{{t_m}}^i\ln \left( {{{\left( {1 - {\lambda ^i}} \right)}^{\sum\limits_{j \ne i} {a_j^i{}^1s_{{t_m}}^j} }}} \right)
  + \sum\limits_m {}^1Y_{{t_m}}^i\ln \left( {1 - {{\left( {1 - {\lambda ^i}} \right)}^{\sum\limits_{j \ne i} {a_j^i{}^1s_{{t_m}}^j} }}} \right)
\end{eqnarray}
where
\begin{eqnarray}
\nonumber
{}^1X_{{t_m}}^i & = & \left({1-{}^1s_{{t_m}}^i}\right)
\left({1-{}^1s_{{t_m} + 1}^i} \right) \ \mbox{and} \\ \nonumber
{}^1Y_{{t_m}}^i & = & \left({1-{}^1s_{{t_m}}^i}\right)
\left({1-{}^2s_{{t_m} + 1}^i} \right){}^1s_{{t_m} + 1}^i.
\end{eqnarray}
The maximum value of $L_1$ cannot be obtained straightforwardly by setting
zero its derivative with respect to $a_j^i$, because $a_j^i$ appears in the
exponential term and the values of $\lambda^i$ are unknown. We resort to
the mean-field approximation to solve this problem (step 2). Specifically,
for node $i$ in the virtual layer, the fraction
$\sum\limits_{j \ne i} {{}^1s_{{t_m}}^ja_j^i}$ of A-neighbors
is approximately equal to the fraction of A-nodes in the whole layer excluding
node $i$ itself:
\begin{eqnarray} \label{eq:virtual_mean_main}
\sum\limits_{j \ne i} {{}^1s_{{t_m}}^ja_j^i}
\approx \frac{{{}^1{k^i}}}{{N - 1}}{}^1\theta _{{t_m}}^i,
\end{eqnarray}
where ${}^1{k^i}$ is the degree of node $i$,
${}^1\theta_{{t_m}}^i=\sum\limits_{j\ne i}{{}^1s_{{t_m}}^j}$ is the number
of A-nodes excluding node $i$ itself. Letting
\begin{displaymath}
{}^1{\gamma^i}={\left({1-{\lambda^i}}\right)^{\frac{{{}^1{k^i}}}{{N-1}}}},
\end{displaymath}
we rewrite Eq.~(\ref{eq:virtual_like_main}) concisely as
\begin{eqnarray} \label{eq:virtual_modif_L1_main}
{\hat L_1}\left( {{}^1{\gamma ^i}} \right) =
\sum\limits_m {\left[ \begin{array}{l}
{}^1X_{{t_m}}^i\ln \left( {{{\left( {{}^1{\gamma ^i}}\right)}^{{}^1\theta_{{t_m}}^i}}} \right)
 + {}^1Y_{{t_m}}^i\ln \left( {1 - {{\left( {{}^1{\gamma ^i}} \right)}^{{}^1\theta _{{t_m}}^i}}} \right)
\end{array} \right]}.
\end{eqnarray}
Differentiating $\hat L_1\left({}^1{\gamma ^i}\right)$ with respect to
${}^1{\gamma ^i}$ and setting it to zero, we get the equation
\begin{equation}
\sum\limits_m {{}^1Y_{{t_m}}^i{}^1\theta_{{t_m}}^i
\frac{{{{\left( {{}^1{\gamma ^i}} \right)}^{{}^1\theta _{{t_m}}^i}}}}
{{1 - {{\left( {{}^1{\gamma ^i}} \right)}^{{}^1\theta _{{t_m}}^i}}}}} =
\sum\limits_m {{}^1X_{{t_m}}^i{}^1\theta _{{t_m}}^i},
\end{equation}
whose solutions give the values of ${}^1{\gamma^i}$ denoted as
${}^1{\gamma^i} = {}^1{\tilde \gamma^i}$.

To transform the MLE problem into a linear system (step 3), we take the
derivative of Eq.~(\ref{eq:virtual_like_main}) with respect to $a^i_l$ and set
it to zero, which gives the following high-dimensional nonlinear equation
(the detailed reconstruction process for the virtual and physical layers is
given in Secs.~IV and V of SI, respectively):
\begin{displaymath} %\label{eq:virtual_deriv_a}
\sum\limits_m {{}^1Y_{{t_m}}^i{}^1s_{{t_m}}^l\frac{{{{\left({1 - {\lambda^i}} \right)}^{\sum\limits_{j \ne i} {a_j^i~{}^1s_{{t_m}}^j} }}}}{{1 - {{\left( {1 - {\lambda ^i}} \right)}^{\sum\limits_{j \ne i} {a_j^i~{}^1s_{{t_m}}^j} }}}}}  = \sum\limits_m {{}^1X_{{t_m}}^i{}^1s_{{t_m}}^l}.
\end{displaymath}
To obtain a solution, we take advantage of Eq.~(\ref{eq:virtual_mean_main}) and
use the first-order approximation. The result is a solvable linear system
expressed in the matrix form
\begin{equation}
\bm{\Lambda}\cdot\bm{\eta} = \bm{\zeta},
\end{equation}
with the vector
\begin{displaymath}
\bm{\eta} = \big[{a_1^i\ln \left({1-{\lambda^i}}\right)},
\ldots,{a_{i-1}^i\ln \left({1-{\lambda^i}} \right)},
{a_{i+1}^i\ln \left( {1 - {\lambda ^i}} \right)},
\ldots, {a_N^i\ln \left( {1 - {\lambda ^i}} \right)}\big]^T
\end{displaymath}
(see Sec.~IV of SI for the detailed forms of $\bm{\Lambda}$ and $\bm{\zeta}$).
With the available time series ${}^1s_{t_{m}}^i$ and ${}^2s_{t_m}^i$, the
values of the elements of the matrix $\bm{\Lambda}$ and the components of
the vector $\bm{\zeta}$ can be calculated, so that the vector $\bm{\eta}$
characterizing the connectivity of node $i$ can be solved. Because
$\ln{(1-\lambda^i)}<0$ is a constant, the value of
$- a_l^i\ln \left( {1 - {\lambda ^i}} \right)$ is much above zero
for $a^i_l=1$ or close to zero for $a^i_l=0$. A threshold value can then
be readily set to distinguish the existent from the nonexistent links: a pair
of nodes $i$ and $l$ are connected in the virtual layer if the value of
$-a_l^i\ln \left({1-{\lambda^i}}\right)$ is larger than the threshold
(see Sec.~VI in SI for the criterion to choose the threshold).

\paragraph*{Evaluation metrics.}
We use three metrics~\cite{LSWGL:2017} to characterize the performance of our
reconstruction framework: the area under the receiver operating characteristic
curve (AUROC), the area under the precision-recall curve (AUPR), and the
Success rate.

To define AUROC and AUPR, it is necessary to calculate three basic quantities:
TPR (true positive rate), FPR (false positive rate), and
Recall~\cite{LSWGL:2017}. In particular, TPR is defined as
\begin{eqnarray}\label{index1_main}
\mbox{TPR}(l)=\frac{\mbox{TP}(l)}{T},
\end{eqnarray}
where $l$ is the cut-off index in the list of the predicted links,
$\mbox{TP}(l)$ is the number of true positives in the top $l$ predictions
in the link list, and $T$ is the number of positives.

FPR is defined as
\begin{eqnarray}\label{index2_main}
\mbox{FPR}(l)=\frac{\mbox{FP}(l)}{Q},
 \end{eqnarray}
where $\mbox{FP}(l)$ is the number of false positives in the top $l$
entries in the predicted link list, and $Q$ is the number of negatives
by the golden standard.

Recall and Precision are defined as
\begin{eqnarray}\label{index3_main}
\mbox{Recall}(l)=\mbox{TPR}(l)=\frac{\mbox{TP}(l)}{T}.
\end{eqnarray}
and
\begin{eqnarray}
\mbox{Precision}(l)=\frac{\mbox{TP}(l)}{\mbox{TP}(l)+\mbox{FP}(l)}
=\frac{\mbox{TP}(l)}{l},
\end{eqnarray}
respectively. Varying the value of $l$ from $0$ to $N$, we plot two sequences
of points: [$\mbox{FPR}(l)$, $\mbox{TPR}(l)$]
and [$\mbox{Recall}(l)$, $\mbox{Precision}(l)$].
The area under the two curves correspond to the values of AUROC and AUPR,
respectively. For perfect reconstruction, we have AUROC=1 and AUPR=1. In
the worst case (completely random), we have AUROC=0.5 and AUPR=$T/2N$.

Let $n_1$ and $n_2$ be the numbers of the actual and nonexistent links in
the network, respectively, $n_3$ and $n_4$ be the numbers of the predicted
existent and nonexistent links. The Success rates for actual links (SREL)
and nonexistent links (SRNL) are defined as $n_3/n_1$ and $n_4/n_2$,
respectively. The normalized Success rate is
$\sqrt{\mbox{SQEL}\times \mbox{SRNL}}$~\cite{SWFDL:2014}.

\acknowledgments

This work was supported by NSFC under Grant Nos.~61473001 and 11331009. YCL
is supported by ONR through Grant No.~N00014-16-1-2828. XL is supported by
the National Science Fund for Distinguished Young Scholars of China
(No.~61425019) and NSFC under Grant No.~71731004.

\onecolumngrid
\newpage

\begin{center}
{\large Supplementary Information for}
\end{center}
\begin{center}
{\large\bf Data based reconstruction of complex multiplex networks}
\end{center}

\begin{center}
Chuang Ma, Han-Shuang Chen, Xiang Li, Ying-Cheng Lai, and Hai-Feng Zhang
\end{center}
%\begin{center}
%Corresponding author: Hai-Feng Zhang (haifengzhang1978@gmail.com)
%\end{center}

{\tableofcontents}

\newpage

\section{UAU-SIS dynamics on multiplex networks with a double-layer
structure} \label{SI:model}
\begin{figure}[h]
\centering
\includegraphics[width=0.8\linewidth]{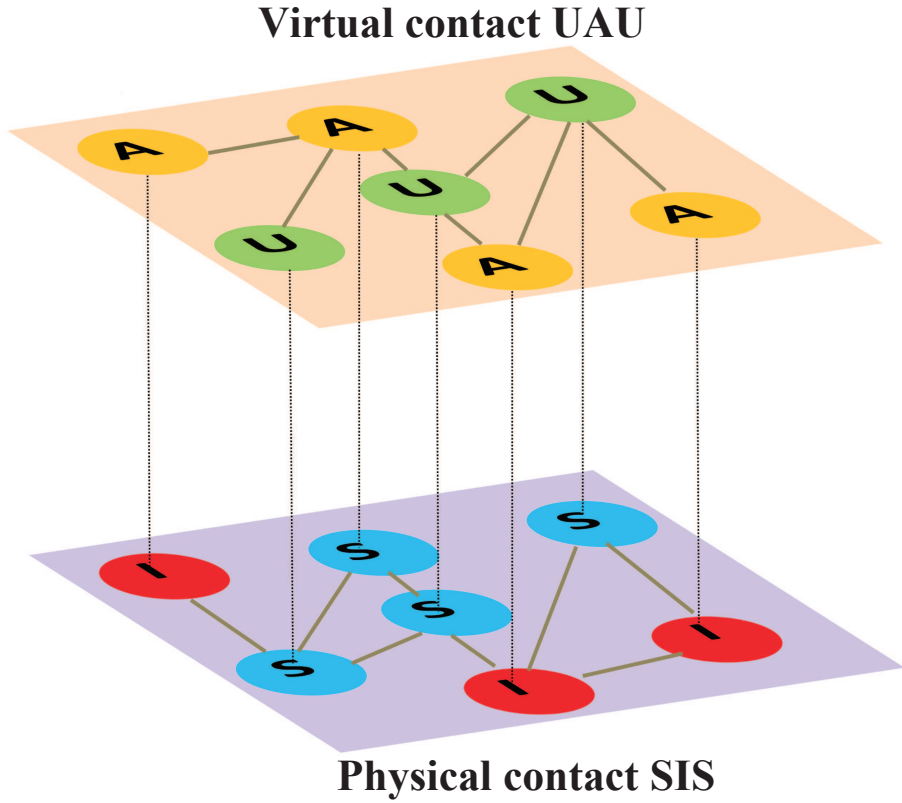}
\caption{ {\bf Schematic illustration a multiplex network with a double layer
interacting structure}. The upper layer (virtual contact) supports awareness
diffusion, where each node has two possible states: unaware (U) or aware (A).
The bottom layer (physical contact) hosts epidemic spreading dynamics, where
a node can be in the susceptible (S) or infected (I) state.}
\label{figS:sketch}
\end{figure}

The UAU-SIS model was originally articulated to describe the
competition between social awareness and disease spreading on double layer
networks, where the physical contact layer supports an epidemic process and
the virtual contact layer supports awareness diffusion~\cite{GGA:2013}. The
two layers share exactly the same set of nodes but their connection patterns
are different. Spreading of awareness in the virtual layer is described by
the conventional SIS dynamics: an unaware node (U) is informed by an aware
neighbor (A) with probability $\lambda$, and an aware node can lose
awareness and returns to the U state with probability $\delta$. Epidemic
dynamics in the physical layer are also of the SIS type, where an infected (I)
node can infect its susceptible (S) neighbors with probability $\beta$, and
an I-node returns to the S state with probability $\mu$. There is interlayer
interaction, which can be described, as follows. An I-node in the physical
layer is automatically aware of the infection and changes to the A state
immediately in the virtual layer. If an S-node is in the A state, its
ability to infect is reduced by a factor $0\leq\gamma<1$. Figure~\ref{figS:sketch}
presents a schematic illustration of the double layer network with the
described interacting dynamical processes.

\section{Transition tree of UAU-SIS model} \label{SI:Tree}

\begin{figure}[h]
\centering
\includegraphics[width=\linewidth]{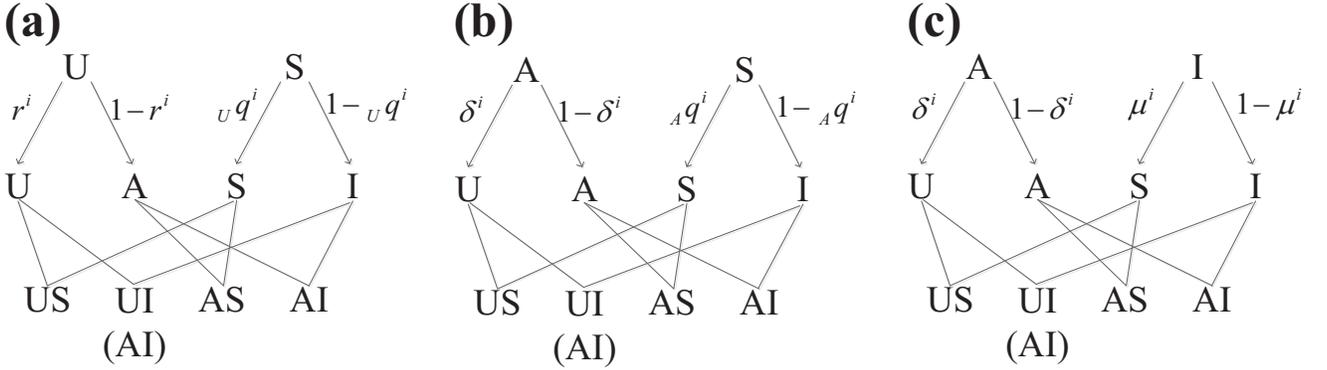}
\caption{ {\bf Transition probability tree of coupled UAU-SIS dynamics
on duplex networks}. The notations are:
AI - aware and infected, UI - unaware and infected (redundant to the AI state),
AS - aware and susceptible, and US - unaware and susceptible.}
\label{fig:trans-tree}
\end{figure}

Let $\sideset{_U}{}{\mathop{\beta}}$ and
$\sideset{_A}{}{\mathop{\beta}}=\gamma\cdot\sideset{_U}{}{\mathop{\beta}}$
be the infection rates of unaware and aware S-node, respectively.
For node $i$, $\sum\limits_{j \ne i} {a_j^i~{}^1s_t^j}$ and
$\sum\limits_{j \ne i} {b_j^i~{}^2s_t^j}$ are the numbers of A-neighbors
in the virtual layer and I-neighbors in the physical layer, respectively.
Three probabilities are needed to describe the network spreading dynamics:
(1) $r_i^t$, the probability that node $i$ is not informed by any neighbor,
(2) ${}_Uq_t^i$, the probability that node $i$ is not infected by any neighbor
if $i$ was unaware, and (3) ${}_Aq_t^i$, the probability that node $i$ is not
infected by any neighbor if $i$ was aware. In the absence of any dynamical
correlation, the three probabilities are given by
\begin{eqnarray} \label{eq:transition}
\begin{array}{l}
r_t^i = {\left( {1 - {\lambda ^i}} \right)^{\sum\limits_{j \ne i} {a_j^i{}^1s_t^j} }},\\
{}_Uq_t^i = {\left( {1 - {}_U{\beta ^i}} \right)^{\sum\limits_{j \ne i} {b_j^i{}^2s_t^j} }},\\
{}_Aq_t^i = {\left( {1 - {}_A{\beta ^i}} \right)^{\sum\limits_{j \ne i} {b_j^i{}^2s_t^j} }}.
\end{array}
\end{eqnarray}
A tacit assumption in Ref.~\cite{GGA:2013} is that diffusion of awareness
in the virtual layer occurs before epidemic spreading in the physical layer.
In our work, we do not require that the two types of spreading dynamics
occur in any particular order. Figure~\ref{fig:trans-tree} presents the
transition probability tree of the UAU-SIS coupling dynamics on the
duplex networks.

From Fig.~\ref{fig:trans-tree} and Eq.~(\ref{eq:transition}), we have that
the probabilities of node $i$ being in the US, AS and AI states at $t+1$ when
it is in the US state at time $t$ are
\begin{eqnarray} \label{eq:US}
\begin{array}{l}
P^{US\rightarrow US} = r_t^i{}_Uq_t^i,\\
P^{US\rightarrow AS} = \left( {1 - r_t^i} \right){}_Uq_t^i,\\
P^{US\rightarrow AI} = r_t^i\left( {1 - {}_Uq_t^i} \right) + \left( {1 - r_t^i} \right)\left( {1 - {}_Uq_t^i} \right) = 1 - {}_Uq_t^i.
\end{array}
\end{eqnarray}
The probabilities of node $i$ being in the US, AS and AI states at $t+1$ when
it is in the AS state at time $t$ are
\begin{eqnarray} \label{eq:US_1}
\begin{array}{l}
P^{AS\rightarrow US} = {\delta ^i}{}_Aq_t^i,\\
P^{AS\rightarrow AS}= \left( {1 - {\delta ^i}} \right){}_Aq_t^i,\\
P^{AS\rightarrow AI} = {\delta ^i}\left( {1 - {}_Aq_t^i} \right) + \left( {1 - {\delta ^i}} \right)\left( {1 - {}_Aq_t^i} \right) = 1 - {}_Aq_t^i.
\end{array}
\end{eqnarray}
The probabilities for node $i$ to be in the US, AS and AI states at $t+1$
if it is in the AI state at time $t$ are
\begin{eqnarray} \label{eq:AI}
\begin{array}{l}
P^{AI\rightarrow US} ={\delta ^i}{\mu ^i}, \\
P^{AI\rightarrow AS}= \left( {1 - {\delta ^i}} \right){\mu ^i}, \\
P^{AI\rightarrow AI} = 1 - {\mu ^i}.
\end{array}
\end{eqnarray}

\section{Likelihood function} \label{SI:Likelihood}

The likelihood function of a node can be written in the following compact form:
\begin{eqnarray} \label{eq:likelihood}
& & \quad P\left( {{{\left\{ {{}^1s_{{t_m} + 1}^i,{}^2s_{{t_m} + 1}^i} \right\}}_{m = 1 \cdots M}}|{{\left\{ {{}^1s_{{t_m}}^j,{}^2s_{{t_m}}^j} \right\}}_{j = 1 \cdots N,m = 1 \cdots M}},{\mathbf{a}^i},{\mathbf{b}^i},{\lambda ^i},{}_U{\beta ^i},{}_A{\beta ^i},{\delta ^i},{\mu ^i}} \right) \\ \nonumber
& = & {\prod\limits_m {\left[ {{{\left( {r_{{t_m}}^i{}_Uq_{{t_m}}^i} \right)}^{\left( {1 - {}^1s_{{t_m} + 1}^i} \right)\left( {1 - {}^2s_{{t_m} + 1}^i} \right)}}{{\left( {\left( {1 - r_{{t_m}}^i} \right){}_Uq_{{t_m}}^i} \right)}^{{}^1s_{{t_m} + 1}^i\left( {1 - {}^2s_{{t_m} + 1}^i} \right)}}{{\left( {1 - {}_Uq_{{t_m}}^i} \right)}^{{}^1s_{{t_m} + 1}^i{}^2s_{{t_m} + 1}^i}}} \right]} ^{\left( {1 - {}^1s_{{t_m}}^i} \right)\left( {1 - {}^2s_{{t_m}}^i} \right)}} \\ \nonumber
& & \quad \quad {\left[ {{{\left( {{\delta ^i}{}_Aq_{{t_m}}^i} \right)}^{\left( {1 - {}^1s_{{t_m} + 1}^i} \right)\left( {1 - {}^2s_{{t_m} + 1}^i} \right)}}{{\left( {\left( {1 - {\delta ^i}} \right){}_Aq_{{t_m}}^i} \right)}^{{}^1s_{{t_m} + 1}^i\left( {1 - {}^2s_{{t_m} + 1}^i} \right)}}{{\left( {1 - {}_Aq_{{t_m}}^i} \right)}^{{}^1s_{{t_m} + 1}^i{}^2s_{{t_m} + 1}^i}}} \right]^{{}^1s_{{t_m}}^i\left( {1 - {}^2s_{{t_m}}^i} \right)}} \\ \nonumber
& & \quad \quad {\left[ {{{\left( {{\delta ^i}{\mu ^i}} \right)}^{\left( {1 - {}^1s_{{t_m} + 1}^i} \right)\left( {1 - {}^2s_{{t_m} + 1}^i} \right)}}{{\left( {\left( {1 - {\delta ^i}} \right){\mu ^i}} \right)}^{{}^1s_{{t_m} + 1}^i\left( {1 - {}^2s_{{t_m} + 1}^i} \right)}}{{\left( {1 - {\mu ^i}} \right)}^{{}^1s_{{t_m} + 1}^i{}^2s_{{t_m} + 1}^i}}} \right]^{{}^1s_{{t_m}}^i{}^2s_{{t_m}}^i}}.
\end{eqnarray}
In the UAU-SIS model, we assume that a node, once it is infected in
the physical layer, enters into the A state immediately in the virtual layer:
${}^2s_{{t}}^i=1$ indicates ${}^1s_{{t}}^i=1$ at any time $t$. As a result,
we have ${}^1s_{{t_m}}^i{}^2s_{{t_m}}^i = {}^2s_{{t_m}}^i$ and
${}^1s_{{t_m} + 1}^i{}^2s_{{t_m} + 1}^i = {}^2s_{{t_m} + 1}^i$. Note that a
node in the U state in the virtual layer cannot be in the I state in the
physical layer, i.e., ${}^1s_{{t}}^i=0$ indicates ${}^2s_{{t}}^i=0$ at any
time $t$, which leads to
\begin{eqnarray}
\nonumber
\left( {1 - {}^1s_{{t_m}}^i} \right)\left( {1 - {}^2s_{{t_m}}^i} \right)
& = & 1 - {}^1s_{{t_m}}^i \\ \nonumber
\left( {1 - {}^1s_{{t_m} + 1}^i}\right)\left({1-{}^2s_{{t_m} + 1}^i}\right)
& = & 1 - {}^1s_{{t_m} + 1}^i.
\end{eqnarray}
Equation~(\ref{eq:likelihood}) appears complicated, but it can be reduced to
some simple form. For example, assuming that node $i$ at $t_m$ is in the US
state: ${}^1s_{{t_m}}^i=0$ and ${}^2s_{{t_m}}^i=0$, we need to retain only
one term in the product:
\begin{displaymath}
{\left[ {{{\left( {r_{{t_m}}^i{}_Uq_{{t_m}}^i} \right)}^{\left( {1 - {}^1s_{{t_m} + 1}^i} \right)\left( {1 - {}^2s_{{t_m} + 1}^i} \right)}}{{\left( {\left( {1 - r_{{t_m}}^i} \right){}_Uq_{{t_m}}^i} \right)}^{{}^1s_{{t_m} + 1}^i\left( {1 - {}^2s_{{t_m} + 1}^i} \right)}}{{\left( {1 - {}_Uq_{{t_m}}^i} \right)}^{{}^1s_{{t_m} + 1}^i{}^2s_{{t_m} + 1}^i}}} \right]},
\end{displaymath}
which can be further reduced to
${\left( {1 - r_{{t_m}}^i} \right){}_Uq_{{t_m}}^i}$ if ${}^1s_{{t_m+1}}^i=1$
and ${}^2s_{{t_m+1}}^i=0$ (i.e., in the AS state at the next time step).
This corresponds to the transition probability from the US state to the AS
state:
\begin{displaymath}
P^{US\rightarrow AS} = \left( {1 - r_t^i} \right){}_Uq_t^i
\end{displaymath}
in Eq.~(\ref{eq:US}). In general, Eq.~(\ref{eq:likelihood}) contains all the
transition probabilities in Eqs.~(\ref{eq:US})-(\ref{eq:AI}).

After some algebra, we obtain the logarithmic form of
Eq.~(\ref{eq:likelihood}) as
\begin{eqnarray} \label{eq:comb}
L\left( {{\mathbf{a^i}},{\mathbf{b^i}},{\lambda ^i},{}_U{\beta ^i},{}_A{\beta ^i},{\delta ^i},{\mu ^i}} \right) =
{L_0}\left( {{\delta ^i},{\mu ^i}} \right)+{L_1}\left( {{\mathbf{a^i}},{\lambda ^i}} \right)+{L_2}\left( {{\mathbf{b^i}},{}_U{\beta ^i},{}_A{\beta ^i}} \right),
\end{eqnarray}
where
\begin{eqnarray} \label{eq:L0}
{L_0}\left( {{\delta ^i},{\mu ^i}} \right) = \sum\limits_m {\left[ \begin{array}{l}
{}^1s_{{t_m}}^i\left( {1 - {}^1s_{{t_m} + 1}^i} \right)\ln \left( {{\delta ^i}} \right) + {}^1s_{{t_m}}^i{}^1s_{{t_m} + 1}^i\left( {1 - {}^2s_{{t_m} + 1}^i} \right)\ln \left( {1 - {\delta ^i}} \right)\\
 + {}^2s_{{t_m}}^i\left( {1 - {}^2s_{{t_m} + 1}^i} \right)\ln \left( {{\mu ^i}} \right) + {}^2s_{{t_m}}^i{}^2s_{{t_m} + 1}^i\ln \left( {1 - {\mu ^i}} \right)
\end{array} \right]}.
\end{eqnarray}
Note that Eq.~(\ref{eq:L0}) does not rely on any information about the
network structure. The quantity that does contain the information is
${L_1}\left( {{a^i},{\lambda ^i}} \right)$, which depends on the connectivity
of node $i$ in the virtual layer. It can be written as
\begin{eqnarray} \label{eq:virtual_like}
{L_1}\left( {{\mathbf{a^i}},{\lambda ^i}} \right) = \sum\limits_m {\left[ {{}^1X_{{t_m}}^i\ln \left( {{{\left( {1 - {\lambda ^i}} \right)}^{\sum\limits_{j \ne i} {a_j^i{}^1s_{{t_m}}^j} }}} \right) + {}^1Y_{{t_m}}^i\ln \left( {1 - {{\left( {1 - {\lambda ^i}} \right)}^{\sum\limits_{j \ne i} {a_j^i{}^1s_{{t_m}}^j} }}} \right)} \right]} ,
\end{eqnarray}
where
\begin{eqnarray}
\nonumber
{}^1X_{{t_m}}^i & = & \left( {1 - {}^1s_{{t_m}}^i} \right)
\left( {1 - {}^1s_{{t_m} + 1}^i} \right) \ \ \mbox{and} \\ \nonumber
{}^1Y_{{t_m}}^i & = & \left( {1 - {}^1s_{{t_m}}^i} \right)
\left( {1 - {}^2s_{{t_m} + 1}^i} \right){}^1s_{{t_m} + 1}^i.
\end{eqnarray}
Similarly, the quantity
${L_2}\left( {{b^i},{}_U{\beta ^i},{}_A{\beta ^i}} \right)$ that depends on
the connectivity of node $i$ in the physical layer is given by
\begin{eqnarray} \label{eq:physical_like}
{L_2}\left( {{\mathbf{b^i}},{}_U{\beta ^i},{}_A{\beta ^i}} \right) =
& & \\ \nonumber
& & \sum\limits_m \left[ {{}_U^2X_{{t_m}}^i\ln \left( {{{\left( {1 - {}_U{\beta ^i}} \right)}^{\sum\limits_{j \ne i} {b_j^i{}^2s_{{t_m}}^j} }}} \right) + {}_U^2Y_{{t_m}}^i\ln\left( {1 - {{\left( {1 - {}_U{\beta ^i}} \right)}^{\sum\limits_{j \ne i} {b_j^i{}^2s_{{t_m}}^j} }}} \right)} \right] \\ \nonumber
& + & \sum\limits_m \left[ {{}_A^2X_{{t_m}}^i\ln \left( {{{\left( {1 - {}_A{\beta ^i}} \right)}^{\sum\limits_{j \ne i} {b_j^i{}^2s_{{t_m}}^j} }}} \right) + {}_A^2Y_{{t_m}}^i\ln \left( {1 - {{\left( {1 - {}_A{\beta ^i}} \right)}^{\sum\limits_{j \ne i} {b_j^i{}^2s_{{t_m}}^j} }}} \right)} \right]
\end{eqnarray}
where
\begin{eqnarray}
\nonumber
{}_U^2X_{{t_m}}^i & = & \left( {1 - {}^1s_{{t_m}}^i} \right)\left( {1 - {}^2s_{{t_m} + 1}^i} \right), \\ \nonumber
{}_U^2Y_{{t_m}}^i & = & \left( {1 - {}^1s_{{t_m}}^i} \right){}^2s_{{t_m} + 1}^i,
\\ \nonumber
{}_A^2X_{{t_m}}^i & = & {}^1s_{{t_m}}^i\left( {1 - {}^2s_{{t_m}}^i} \right)\left( {1 - {}^2s_{{t_m} + 1}^i} \right), \ \ \mbox{and} \\ \nonumber
{}_A^2Y_{{t_m}}^i & = & {}^1s_{{t_m}}^i\left( {1 - {}^2s_{{t_m}}^i} \right){}^2s_{{t_m} + 1}^i.
\end{eqnarray}
In principle, Eq.~(\ref{eq:comb}) indicates that one can maximize $L_1$ and
$L_2$ with respect to $a^i_j$ and $b^i_j$, respectively, to uncover the
connectivity of node $i$. However, the conventional maximization process
leads to equations that cannot be solved because the quantity $a_j^i$ ($b_j^i$)
appears in the exponential term and the values of $\lambda^i$ (or
${}_U{\beta ^i}$) are unknown. We exploit the mean-field approximation to
maximize $L_1$ and $L_2$.

\section{Reconstruction of virtual layer} \label{SI:Recon_virtual}

To infer the neighbors of node $i$ in the virtual layer, we impose the
mean-field approximation on $L_1$:
\begin{eqnarray} \label{eq:virtual_mean}
\sum\limits_{j \ne i} {{}^1s_{{t_m}}^ja_j^i} \approx
\frac{{{}^1{k^i}}}{{N - 1}}{}^1\theta _{{t_m}}^i,
\end{eqnarray}
where $N$ and ${}^1{k^i}$ are the number of nodes and the degree of node
$i$ in the virtual layer, respectively, and the number of A-nodes in the
virtual layer (excluding node $i$ itself) is
${}^1\theta _{{t_m}}^i = \sum\limits_{j \ne i} {{}^1s_{{t_m}}^j}$. A new
unknown parameter ${}^1{k^i}$ emerges in Eq.~(\ref{eq:virtual_like}) when we
substitute Eq.~(\ref{eq:virtual_mean}) into Eq.~(\ref{eq:virtual_like}).
To simplify the analysis, we let
\begin{eqnarray} \label{eq:trick}
{}^1{\gamma ^i} = {\left( {1 - {\lambda ^i}} \right)^{\frac{{{}^1{k^i}}}{{N - 1}}}},
\end{eqnarray}
to obtain
\begin{displaymath}
{\left( {1 - {\lambda ^i}} \right)^{\sum\limits_{j \ne i} {a_j^i~{}^1s_{{t_m}}^j} }} = {\left( {1 - {\lambda ^i}} \right)^{\frac{{{}^1{k^i}}}{{N - 1}}{}^1\theta_{{t_m}}^i}} = {\left( {{}^1{\gamma ^i}} \right)^{{}^1\theta _{{t_m}}^i}}.
\end{displaymath}
Equation~(\ref{eq:virtual_like}) can then be written concisely as
\begin{eqnarray} \label{eq:virtual_modif_L1}
{\hat L_1}\left( {{}^1{\gamma ^i}} \right) = \sum\limits_m
{\left[ \begin{array}{l}
{}^1X_{{t_m}}^i\ln \left( {{{\left( {{}^1{\gamma ^i}} \right)}^{{}^1\theta _{{t_m}}^i}}} \right)
 + {}^1Y_{{t_m}}^i\ln \left( {1 - {{\left( {{}^1{\gamma ^i}} \right)}^{{}^1\theta _{{t_m}}^i}}} \right)
\end{array} \right]} .
\end{eqnarray}
Differentiating $\hat L_1\left( {}^1{\gamma ^i} \right)$ with respect to
${}^1{\gamma ^i}$ and setting it to zero, we get
\begin{displaymath}
\sum\limits_m {{}^1Y_{{t_m}}^i{}^1\theta _{{t_m}}^i\frac{{{{\left( {{}^1{\gamma ^i}} \right)}^{{}^1\theta _{{t_m}}^i}}}}{{1 - {{\left( {{}^1{\gamma ^i}} \right)}^{{}^1\theta _{{t_m}}^i}}}}} = \sum\limits_m {{}^1X_{{t_m}}^i{}^1\theta_{{t_m}}^i},
\end{displaymath}
leading to ${}^1{\gamma ^i} = {}^1{\tilde \gamma ^i}$.

Treating $a^i_l$ as a continuous variance, we can further differentiate
Eq.~(\ref{eq:virtual_like}) with respect to $a^i_l$ and set it to zero,
which gives
\begin{eqnarray} \label{eq:virtual_deriv_a}
\sum\limits_m {{}^1Y_{{t_m}}^i~{}^1s_{{t_m}}^l\frac{{{{\left( {1 - {\lambda ^i}} \right)}^{\sum\limits_{j \ne i} {a_j^i~{}^1s_{{t_m}}^j} }}}}{{1 - {{\left( {1 - {\lambda ^i}} \right)}^{\sum\limits_{j \ne i} {a_j^i{}^1s_{{t_m}}^j} }}}}}  = \sum\limits_m {{}^1X_{{t_m}}^i{}^1s_{{t_m}}^l}.
\end{eqnarray}
To obtain analytical solutions of Eq.~(\ref{eq:virtual_deriv_a}) is
not feasible due to its nonlinear and high-dimensional nature. We thus
resort to the first-order Taylor expansion. In particular, we expand
$a^x/(1 - a^x)$ in the limit $x\rightarrow x_0$ to obtain
\begin{eqnarray} \label{eq:taylor}
\frac{{{a^x}}}{{1 - {a^x}}} \approx \frac{{{a^{{x_0}}}}}{{1 - {a^{{x_0}}}}} + \frac{{{a^{{x_0}}}\ln a}}{{{{\left( {1 - {a^{{x_0}}}} \right)}^2}}}\left( {x - {x_0}} \right) = \frac{{{a^{{x_0}}}}}{{1 - {a^{{x_0}}}}} - \frac{{{a^{{x_0}}}\ln {a^{{x_0}}}}}{{{{\left( {1 - {a^{{x_0}}}} \right)}^2}}} + \frac{{{a^{{x_0}}}\ln a}}{{{{\left( {1 - {a^{{x_0}}}} \right)}^2}}}x.
\end{eqnarray}
Setting $x = \sum\limits_{j \ne i} {a_j^i~{}^1s_{{t_m}}^j} $,
$a= 1 - {\lambda ^i}$, and
${x_0} = \frac{{{}^1{k^i}}}{{N - 1}}{}^1\theta _{{t_m}}^i$
[Eq.~(\ref{eq:virtual_mean}) implies that $x\approx x_0$], we have
\begin{displaymath}
{a^{{x_0}}}={\left({{}^1{{\tilde\gamma }^i}}\right)^{{}^1\theta_{{t_m}}^i}}
\end{displaymath}
from Eq.~(\ref{eq:trick}). That is, the term
\begin{displaymath}
{\frac{{{{\left( {1 - {\lambda ^i}} \right)}^{\sum\limits_{j \ne i} {a_j^i~{}^1s_{{t_m}}^j} }}}}{{1 - {{\left( {1 - {\lambda ^i}} \right)}^{\sum\limits_{j \ne i} {a_j^i~{}^1s_{{t_m}}^j} }}}}}
\end{displaymath}
in Eq.~(\ref{eq:virtual_deriv_a}) can be expressed as a linear form based
on Eq.~(\ref{eq:virtual_mean}.)

Collecting all the approximations, we transform Eq.~(\ref{eq:virtual_deriv_a})
into a solvable linear system as
\begin{eqnarray} \label{eq:virtual_lin_eq}
\begin{array}{l}
\sum\limits_m {{}^1Y_{{t_m}}^i{}^1G_{{t_m}}^i{}^1s_{{t_m}}^l\ln \left( {1 - {\lambda ^i}} \right)\sum\limits_{j \ne i} {a_j^i~{}^1s_{{t_m}}^j} }
 = \sum\limits_m {\left( {{}^1X_{{t_m}}^i - {}^1Y_{{t_m}}^i{}^1F_{{t_m}}^i} \right)~{}^1s_{{t_m}}^l},
\end{array}
\end{eqnarray}
where
\begin{eqnarray}
\nonumber
{}^1F_{{t_m}}^i & = & \frac{{{{\left( {{}^1{{\tilde \gamma }^i}} \right)}^{{}^1\theta _{{t_m}}^i}}}}{{1 - {{\left( {{}^1{{\tilde \gamma }^i}} \right)}^{{}^1\theta _{{t_m}}^i}}}} - \frac{{{{\left( {{}^1{{\tilde \gamma }^i}} \right)}^{{}^1\theta _{{t_m}}^i}}}}{{{{\left( {1 - {{\left( {{}^1{{\tilde \gamma }^i}} \right)}^{{}^1\theta _{{t_m}}^i}}} \right)}^2}}}{}^1\theta _{{t_m}}^i\ln {}^1{\tilde \gamma ^i} \ \ \mbox{and} \\ \nonumber
{}^1G_{{t_m}}^i & = & \frac{{{{\left( {{}^1{{\tilde \gamma }^i}} \right)}^{{}^1\theta _{{t_m}}^i}}}}{{{{\left( {1 - {{\left( {{}^1{{\tilde \gamma }^i}} \right)}^{{}^1\theta _{{t_m}}^i}}} \right)}^2}}}.
\end{eqnarray}
Setting ${}^1\Phi _{{t_m}}^i = {}^1Y_{{t_m}}^i{}^1G_{{t_m}}^i$ and
${}^1\Gamma _{{t_m}}^i$ = ${}^1X_{{t_m}}^i - {}^1Y_{{t_m}}^i{}^1F_{{t_m}}^i$,
we express Eq.~(\ref{eq:virtual_lin_eq}) as
\begin{eqnarray} \label{eq:virtual_matrix}
\nonumber
& & \left[ {\begin{array}{*{20}{c}}
{\sum\limits_m {{}^1\Phi _{{t_m}}^i{}^1s_{{t_m}}^1{}^1s_{{t_m}}^1} }& \cdots &{\sum\limits_m {{}^1\Phi _{{t_m}}^i{}^1s_{{t_m}}^1{}^1s_{{t_m}}^{i - 1}} }&{\sum\limits_m {{}^1\Phi _{{t_m}}^i{}^1s_{{t_m}}^1{}^1s_{{t_m}}^{i + 1}} }& \cdots &{\sum\limits_m {{}^1\Phi _{{t_m}}^i{}^1s_{{t_m}}^1{}^1s_{{t_m}}^N} }\\
 \vdots &{}& \vdots & \vdots &{}& \vdots \\
{\sum\limits_m {{}^1\Phi _{{t_m}}^i{}^1s_{{t_m}}^{i - 1}{}^1s_{{t_m}}^1} }& \cdots &{\sum\limits_m {{}^1\Phi _{{t_m}}^i{}^1s_{{t_m}}^{i - 1}{}^1s_{{t_m}}^{i - 1}} }&{\sum\limits_m {{}^1\Phi _{{t_m}}^i{}^1s_{{t_m}}^{i - 1}{}^1s_{{t_m}}^{i + 1}} }& \cdots &{\sum\limits_m {{}^1\Phi _{{t_m}}^i{}^1s_{{t_m}}^{i - 1}{}^1s_{{t_m}}^N} }\\
{\sum\limits_m {{}^1\Phi _{{t_m}}^i{}^1s_{{t_m}}^{i + 1}{}^1s_{{t_m}}^1} }& \cdots &{\sum\limits_m {{}^1\Phi _{{t_m}}^i{}^1s_{{t_m}}^{i + 1}{}^1s_{{t_m}}^{i - 1}} }&{\sum\limits_m {{}^1\Phi _{{t_m}}^i{}^1s_{{t_m}}^{i + 1}{}^1s_{{t_m}}^{i + 1}} }& \cdots &{\sum\limits_m {{}^1\Phi _{{t_m}}^i{}^1s_{{t_m}}^{i + 1}{}^1s_{{t_m}}^N} }\\
 \vdots &{}& \vdots & \vdots &{}& \vdots \\
{\sum\limits_m {{}^1\Phi _{{t_m}}^i{}^1s_{{t_m}}^N{}^1s_{{t_m}}^1} }& \cdots &{\sum\limits_m {{}^1\Phi _{{t_m}}^i{}^1s_{{t_m}}^N{}^1s_{{t_m}}^{i - 1}} }&{\sum\limits_m {{}^1\Phi _{{t_m}}^i{}^1s_{{t_m}}^N{}^1s_{{t_m}}^{i + 1}} }& \cdots &{\sum\limits_m {{}^1\Phi _{{t_m}}^i{}^1s_{{t_m}}^N{}^1s_{{t_m}}^N} }
\end{array}} \right] \\
& \times & \left[ {\begin{array}{*{20}{c}}
{a_1^i\ln \left( {1 - {\lambda ^i}} \right)}\\
 \vdots \\
{a_{i - 1}^i\ln \left( {1 - {\lambda ^i}} \right)}\\
{a_{i + 1}^i\ln \left( {1 - {\lambda ^i}} \right)}\\
 \vdots \\
{a_N^i\ln \left( {1 - {\lambda ^i}} \right)}
\end{array}} \right] = \left[ {\begin{array}{*{20}{c}}
{\sum\limits_m {{}^1\Gamma _{{t_m}}^i{}^1s_{{t_m}}^1} }\\
 \vdots \\
{\sum\limits_m {{}^1\Gamma _{{t_m}}^i{}^1s_{{t_m}}^{i - 1}} }\\
{\sum\limits_m {{}^1\Gamma _{{t_m}}^i{}^1s_{{t_m}}^{i + 1}} }\\
 \vdots \\
{\sum\limits_m {{}^1\Gamma _{{t_m}}^i{}^1s_{{t_m}}^N} }
\end{array}} \right].
\end{eqnarray}
The matrix on the left side (labeled as $\bm{\Lambda}$) and the vector (labeled as $\mathbf{\bm{\zeta}}$)
on the right side of Eq.~(\ref{eq:virtual_matrix}) can be calculated from
the time series of the nodal states. The vector
\begin{displaymath}
\bm{\eta}=\big[{a_1^i\ln \left( {1 - {\lambda ^i}} \right)}, \ldots, {a_{i-1}^i\ln \left( {1 - {\lambda ^i}} \right)}, {a_{i+1}^i\ln \left( {1 - {\lambda ^i}} \right)}, \ldots, {a_N^i\ln \left( {1 - {\lambda ^i}} \right)}\big]^T
\end{displaymath}
can then be solved, where $T$ denotes transpose. Note that the quantity
$\ln \left( {1 - {\lambda ^i}} \right)<0$ is a constant even though
$\lambda^i$ is not given, implying that the value of
$- a_j^i\ln \left( {1 - {\lambda ^i}} \right)$ is positively large
for $a^i_j=1$ and near zero for $a^i_j=0$. As a result, the neighbors of
node $i$ in the virtual layer can be ascertained through the solution of
the vector $\bm{\eta}$.

\section{Reconstruction of physical layer} \label{SI:Recon_physical}

The mean-field approximation for the physical layer is
\begin{eqnarray} \label{eq:physical_mean}
\sum\limits_{j \ne i} {{}^2s_{{t_m}}^jb_j^i}\approx \frac{{{}^2{k^i}}}{{N - 1}}{}^2\theta _{{t_m}}^i,
\end{eqnarray}
where ${}^2{k^i}$ is the degree of node $i$ and
${}^2\theta _{{t_m}}^i = \sum\limits_{j \ne i} {{}^2s_{{t_m}}^j} $ is the
number of I-nodes in the physical layer (excluding node $i$ itself). We set
\begin{eqnarray} \label{eq:physical_trick}
\begin{array}{l}
{}_U^2{\gamma ^i} = {\left( {1 - {}_U{\beta ^i}} \right)^{\frac{{{}^2{k^i}}}{{N - 1}}}} \ \mbox{and} \\
{}_A^2{\gamma ^i} = {\left( {1 - {}_A{\beta ^i}} \right)^{\frac{{{}^2{k^i}}}{{N - 1}}}}
\end{array}
\end{eqnarray}
and write Eq.~(\ref{eq:physical_like}) concisely as
\begin{eqnarray} \label{eq:physical_concise}
{\hat L_2}\left( {{}_U^2{\gamma ^i},{}_A^2{\gamma ^i}} \right) = \sum\limits_m {\left\{ \begin{array}{l}
\left[ {{}_U^2X_{{t_m}}^i\ln \left( {{{\left( {{}_U^2{\gamma ^i}} \right)}^{~{}^2\theta _{{t_m}}^i}}} \right) + {}_U^2Y_{{t_m}}^i\ln\left( {1 - {{\left( {{}_U^2{\gamma ^i}} \right)}^{~{}^2\theta _{{t_m}}^i}}} \right)} \right]\\
 + \left[ {{}_A^2X_{{t_m}}^i\ln \left( {{{\left( {{}_A^2{\gamma ^i}} \right)}^{~{}^2\theta _{{t_m}}^i}}} \right) + {}_A^2Y_{{t_m}}^i\ln \left( {1 - {{\left( {{}_A^2{\gamma ^i}} \right)}^{~{}^2\theta _{{t_m}}^i}}} \right)} \right]
\end{array} \right\}}.
\end{eqnarray}
Taking the derivatives of $\hat L_2$ with respect to ${}_U^2{\gamma^i}$ and
${}_A^2{\gamma^i}$ and setting them to zero, we get
\begin{eqnarray} \label{eq:physical_gamma1}
\sum\limits_m {{}_U^2Y_{{t_m}}^i~{}^2\theta _{{t_m}}^i\frac{{{{\left( {{}_U^2{\gamma ^i}} \right)}^{~{}^2\theta _{{t_m}}^i}}}}{{1 - {{\left( {{}_U^2{\gamma ^i}} \right)}^{~{}^2\theta _{{t_m}}^i}}}}}  = \sum\limits_m {{}_U^2X_{{t_m}}^i~{}^2\theta _{{t_m}}^i}
\end{eqnarray}
and
\begin{eqnarray} \label{eq:physical_gamma2}
\sum\limits_m {{}_A^2Y_{{t_m}}^i~{}^2\theta _{{t_m}}^i\frac{{{{\left( {{}_A^2{\gamma ^i}} \right)}^{~{}^2\theta _{{t_m}}^i}}}}{{1 - {{\left( {{}_A^2{\gamma ^i}} \right)}^{{}^2\theta _{{t_m}}^i}}}}}  = \sum\limits_m {{}_A^2X_{{t_m}}^i~{}^2\theta _{{t_m}}^i}\emph{},
\end{eqnarray}
which give
\begin{eqnarray}
\nonumber
{}_U^2{\gamma ^i} & = & {}_U^2{\tilde \gamma ^i} \ \mbox{and} \\ \nonumber
{}_A^2{\gamma ^i} & = & {}_A^2{\tilde \gamma ^i}.
\end{eqnarray}
Similar to the mean-field analysis of the virtual layer, we differentiate
Eq.~(\ref{eq:physical_like}) with respect to $b_l^i$ and set it to zero:
\begin{eqnarray}
\nonumber
\frac{{\partial {L_2}}}{{\partial b_l^i}} & = &
\sum\limits_m
\left[ {\ln \left( {1 - {}_U{\beta ^i}} \right){}_U^2X_{{t_m}}^i~{}^2s_{{t_m}}^l - \ln \left( {1 - {}_U{\beta ^i}} \right){}_U^2Y_{{t_m}}^i~{}^2s_{{t_m}}^l\frac{{{{\left( {1 - {}_U{\beta ^i}} \right)}^{\sum\limits_{j \ne i} {b_j^i~{}^2s_{{t_m}}^j} }}}}{{1 - {{\left( {1 - {}_U{\beta ^i}} \right)}^{\sum\limits_{j \ne i} {b_j^i~{}^2s_{{t_m}}^j} }}}}} \right] \\ \nonumber
& + & \sum\limits_m \left[ {\ln \left( {1 - {}_A{\beta ^i}} \right){}_A^2X_{{t_m}}^i~{}^2s_{{t_m}}^l - \ln \left( {1 - {}_A{\beta ^i}} \right){}_A^2Y_{{t_m}}^i~{}^2s_{{t_m}}^l\frac{{{{\left( {1 - {}_A{\beta ^i}} \right)}^{\sum\limits_{j \ne i} {b_j^i~{}^2s_{{t_m}}^j} }}}}{{1 - {{\left( {1 - {}_A{\beta ^i}} \right)}^{\sum\limits_{j \ne i} {b_j^i~{}^2s_{{t_m}}^j} }}}}} \right] = 0,
\end{eqnarray}
which gives
\begin{eqnarray} \label{eq:physical_b2}
\nonumber
& & \sum\limits_m {\left\{ {\ln \left( {1 - {}_U{\beta ^i}} \right)~{}^2s_{{t_m}}^l{}_U^2Y_{{t_m}}^i\frac{{{{\left( {1 - {}_U{\beta ^i}} \right)}^{\sum\limits_{j \ne i} {b_j^i~{}^2s_{{t_m}}^j} }}}}{{1 - {{\left( {1 - {}_U{\beta ^i}} \right)}^{\sum\limits_{j \ne i} {b_j^i~{}^2s_{{t_m}}^j} }}}} + \ln \left( {1 - {}_A{\beta ^i}} \right){}^2s_{{t_m}}^l{}_A^2Y_{{t_m}}^i\frac{{{{\left( {1 - {}_A{\beta ^i}} \right)}^{\sum\limits_{j \ne i} {b_j^i{}^2s_{{t_m}}^j} }}}}{{1 - {{\left( {1 - {}_A{\beta ^i}} \right)}^{\sum\limits_{j \ne i} {b_j^i~{}^2s_{{t_m}}^j} }}}}} \right\}} \\
& = & \sum\limits_m {\left\{ {\ln \left( {1 - {}_U{\beta ^i}} \right){}_U^2X_{{t_m}}^i~{}^2s_{{t_m}}^l + \ln \left( {1 - {}_A{\beta ^i}} \right){}_A^2X_{{t_m}}^i{}^2s_{{t_m}}^l} \right\}}.
\end{eqnarray}
Setting
\begin{displaymath}
\rho  = \frac{{\ln {}_U^2{{\tilde \gamma }^i}}}{{\ln {}_A^2{{\tilde \gamma }^i}}} = \frac{{\ln \left( {1 - {}_U{\beta ^i}} \right)}}{{\ln \left( {1 - {}_A{\beta ^i}} \right)}},
\end{displaymath}
we can further simplify Eq.~(\ref{eq:physical_b2}) as
\begin{eqnarray} \label{eq:physical_b3}
\begin{array}{l}
\sum\limits_m {\left[ {\rho ~~ {{}^2s_{{t_m}}^l}~ {}_U^2Y_{{t_m}}^i\frac{{{{\left( {1 - {}_U{\beta ^i}} \right)}^{\sum\limits_{j \ne i} {b_j^i~{}^2s_{{t_m}}^j} }}}}{{1 - {{\left( {1 - {}_U{\beta ^i}} \right)}^{\sum\limits_{j \ne i} {b_j^i{}^2s_{{t_m}}^j} }}}} + {}^2s_{{t_m}}^l{}_A^2Y_{{t_m}}^i\frac{{{{\left( {1 - {}_A{\beta ^i}} \right)}^{\sum\limits_{j \ne i} {b_j^i{}^2s_{{t_m}}^j} }}}}{{1 - {{\left( {1 - {}_A{\beta ^i}} \right)}^{\sum\limits_{j \ne i} {b_j^i{}^2s_{{t_m}}^j} }}}}} \right]} \\
 = \sum\limits_m {\left[ {\rho ~~{}_U^2X_{{t_m}}^i{}^2s_{{t_m}}^l + {}_A^2X_{{t_m}}^i{}^2s_{{t_m}}^l} \right]}
\end{array}
\end{eqnarray}
Using Eq.~(\ref{eq:taylor}) and setting
$x = \sum\limits_{j \ne i} {b_j^i~{}^2s_{{t_m}}^j} $ and
${x_0} = \frac{{{}^2{k^i}}}{{N - 1}}{}^2\theta _{{t_m}}^i$, we obtain the
following Taylor expansion:
\begin{eqnarray} \label{eq:physical_taylor1}
\frac{{{{\left( {1 - {}_U{\beta ^i}} \right)}^{\sum\limits_{j \ne i} {b_j^i~{}^2s_{{t_m}}^j} }}}}{{1 - {{\left( {1 - {}_U{\beta ^i}} \right)}^{\sum\limits_{j \ne i} {b_j^i~{}^2s_{{t_m}}^j} }}}} = {}_U^2F_{{t_m}}^i + {}_U^2G_{{t_m}}^i\ln \left( {1 - {}_U{\beta ^i}} \right)\sum\limits_{j \ne i} {b_j^i~{}^2s_{{t_m}}^j},
\end{eqnarray}
where
\begin{eqnarray}
\nonumber
{}_U^2F_{{t_m}}^i & = & \frac{{{{\left( {{}_U^2{{\tilde \gamma }^i}} \right)}^{~{}^2\theta _{{t_m}}^i}}}}{{1 - {{\left( {{}_U^2{{\tilde \gamma }^i}} \right)}^{~{}^2\theta _{{t_m}}^i}}}} - \frac{{{{\left( {{}_U^2{{\tilde \gamma }^i}} \right)}^{~{}^2\theta _{{t_m}}^i}}}}{{{{\left( {1 - {{\left( {{}_U^2{{\tilde \gamma }^i}} \right)}^{~{}^2\theta _{{t_m}}^i}}} \right)}^2}}}~{}^2\theta _{{t_m}}^i\ln {}_U^2{\tilde \gamma ^i} \ \mbox{and} \\ \nonumber
{}_U^2G_{{t_m}}^i & = & \frac{{{{\left( {{}_U^2{{\tilde \gamma }^i}} \right)}^{~{}^2\theta _{{t_m}}^i}}}}{{{{\left( {1 - {{\left( {{}_U^2{{\tilde \gamma }^i}} \right)}^{~{}^2\theta _{{t_m}}^i}}} \right)}^2}}}.
\end{eqnarray}
We also have
\begin{eqnarray} \label{eq:physical_taylor2}
\frac{{{{\left( {1 - {}_A{\beta ^i}} \right)}^{\sum\limits_{j \ne i} {b_j^i~{}^2s_{{t_m}}^j} }}}}{{1 - {{\left( {1 - {}_A{\beta ^i}} \right)}^{\sum\limits_{j \ne i} {b_j^i~{}^2s_{{t_m}}^j} }}}} = {}_A^2F_{{t_m}}^i + {}_A^2G_{{t_m}}^i\ln \left( {1 - {}_A{\beta ^i}} \right)\sum\limits_{j \ne i} {b_j^i~{}^2s_{{t_m}}^j},
\end{eqnarray}
where
\begin{eqnarray}
\nonumber
{}_A^2F_{{t_m}}^i & = & \frac{{{{\left( {{}_A^2{{\tilde \gamma }^i}} \right)}^{~{}^2\theta _{{t_m}}^i}}}}{{1 - {{\left( {{}_A^2{{\tilde \gamma }^i}} \right)}^{~{}^2\theta _{{t_m}}^i}}}} - \frac{{{{\left( {{}_A^2{{\tilde \gamma }^i}} \right)}^{~{}^2\theta _{{t_m}}^i}}}}{{{{\left( {1 - {{\left( {{}_A^2{{\tilde \gamma }^i}} \right)}^{~{}^2\theta _{{t_m}}^i}}} \right)}^2}}}{}^2\theta _{{t_m}}^i\ln {}_A^2{\tilde \gamma ^i} \ \mbox{and} \\ \nonumber
{}_A^2G_{{t_m}}^i & = & \frac{{{{\left( {{}_A^2{{\tilde \gamma }^i}} \right)}^{~{}^2\theta _{{t_m}}^i}}}}{{{{\left( {1 - {{\left( {{}_A^2{{\tilde \gamma }^i}} \right)}^{~{}^2\theta _{{t_m}}^i}}} \right)}^2}}}.
\end{eqnarray}
With these approximations, we can transform Eq.~(\ref{eq:physical_b3})
into the following linear system:
\begin{eqnarray} \label{eq:physical_eq}
\begin{array}{l}
\sum\limits_m {\left[ {\left( {{\rho ^2}~{}_U^2Y_{{t_m}}^i~{}_U^2G_{{t_m}}^i + {}_A^2Y_{{t_m}}^i~{}_A^2G_{{t_m}}^i} \right)~{}^2s_{{t_m}}^l\ln \left( {1 - {}_A{\beta ^i}} \right)\sum\limits_{j \ne i} {b_j^i~{}^2s_{{t_m}}^j} } \right]} \\
 = \sum\limits_m {\left( {\rho ~{}_U^2X_{{t_m}}^i + {}_A^2X_{{t_m}}^i - \rho~{}_U^2Y_{{t_m}}^i~{}_U^2F_{{t_m}}^i - {}_A^2Y_{{t_m}}^i~{}_A^2F_{{t_m}}^i} \right)}~ {}^2s_{{t_m}}^l,
\end{array}
\end{eqnarray}
where
\begin{eqnarray}
\nonumber
{}^2\Phi _{{t_m}}^i & = & {\rho ^2}~{}_U^2Y_{{t_m}}^i~{}_U^2G_{{t_m}}^i + {}_A^2Y_{{t_m}}^i~{}_A^2G_{{t_m}}^i \ \mbox{and} \\ \nonumber
{}^2\Gamma_{{t_m}}^i & = & \rho ~{}_U^2X_{{t_m}}^i + {}_A^2X_{{t_m}}^i - \rho ~{}_U^2Y_{{t_m}}^i~{}_U^2F_{{t_m}}^i - {}_A^2Y_{{t_m}}^i~{}_A^2F_{{t_m}}^i.
\end{eqnarray}
We rewrite Eq.~(\ref{eq:physical_eq}) in the following matrix form:
\begin{eqnarray} \label{eq:physical_matrix}
\nonumber
& & \left[ {\begin{array}{*{20}{c}}
{\sum\limits_m {{}^2\Phi _{{t_m}}^i{}^2s_{{t_m}}^1{}^2s_{{t_m}}^1} }& \cdots &{\sum\limits_m {{}^2\Phi _{{t_m}}^i{}^2s_{{t_m}}^1{}^2s_{{t_m}}^{i - 1}} }&{\sum\limits_m {{}^2\Phi _{{t_m}}^i{}^2s_{{t_m}}^1{}^2s_{{t_m}}^{i + 1}} }& \cdots &{\sum\limits_m {{}^2\Phi _{{t_m}}^i{}^2s_{{t_m}}^1{}^2s_{{t_m}}^N} }\\
 \vdots &{}& \vdots & \vdots &{}& \vdots \\
{\sum\limits_m {{}^2\Phi _{{t_m}}^i{}^2s_{{t_m}}^{i - 1}{}^2s_{{t_m}}^1} }& \cdots &{\sum\limits_m {{}^2\Phi _{{t_m}}^i{}^2s_{{t_m}}^{i - 1}{}^2s_{{t_m}}^{i - 1}} }&{\sum\limits_m {{}^2\Phi _{{t_m}}^i{}^2s_{{t_m}}^{i - 1}{}^2s_{{t_m}}^{i + 1}} }& \cdots &{\sum\limits_m {{}^2\Phi _{{t_m}}^i{}^2s_{{t_m}}^{i - 1}{}^2s_{{t_m}}^N} }\\
{\sum\limits_m {{}^2\Phi _{{t_m}}^i{}^2s_{{t_m}}^{i + 1}{}^2s_{{t_m}}^1} }& \cdots &{\sum\limits_m {{}^2\Phi _{{t_m}}^i{}^2s_{{t_m}}^{i + 1}{}^2s_{{t_m}}^{i - 1}} }&{\sum\limits_m {{}^2\Phi _{{t_m}}^i{}^2s_{{t_m}}^{i + 1}{}^2s_{{t_m}}^{i + 1}} }& \cdots &{\sum\limits_m {{}^2\Phi _{{t_m}}^i{}^2s_{{t_m}}^{i + 1}{}^2s_{{t_m}}^N} }\\
 \vdots &{}& \vdots & \vdots &{}& \vdots \\
{\sum\limits_m {{}^2\Phi _{{t_m}}^i{}^2s_{{t_m}}^N{}^2s_{{t_m}}^1} }& \cdots &{\sum\limits_m {{}^2\Phi _{{t_m}}^i{}^2s_{{t_m}}^N{}^2s_{{t_m}}^{i - 1}} }&{\sum\limits_m {{}^2\Phi _{{t_m}}^i{}^2s_{{t_m}}^N{}^2s_{{t_m}}^{i + 1}} }& \cdots &{\sum\limits_m {{}^2\Phi _{{t_m}}^i{}^2s_{{t_m}}^N{}^2s_{{t_m}}^N} }
\end{array}} \right] \\
& \times & \left[ {\begin{array}{*{20}{c}}
{b_1^i\ln \left( {1 - {}_A{\beta ^i}} \right)}\\
 \vdots \\
{b_{i - 1}^i\ln \left( {1 - {}_A{\beta ^i}} \right)}\\
{b_{i + 1}^i\ln \left( {1 - {}_A{\beta ^i}} \right)}\\
 \vdots \\
{b_N^i\ln \left( {1 - {}_A{\beta ^i}} \right)}
\end{array}} \right] = \left[ {\begin{array}{*{20}{c}}
{\sum\limits_m {{}^2\Gamma _{{t_m}}^i{}^2s_{{t_m}}^1} }\\
 \vdots \\
{\sum\limits_m {{}^2\Gamma _{{t_m}}^i{}^2s_{{t_m}}^{i - 1}} }\\
{\sum\limits_m {{}^2\Gamma _{{t_m}}^i{}^2s_{{t_m}}^{i + 1}} }\\
 \vdots \\
{\sum\limits_m {{}^2\Gamma _{{t_m}}^i{}^2s_{{t_m}}^N} }
\end{array}} \right].
\end{eqnarray}
The matrix on the left side and the vector on the right side can be obtained
from time series. Solution of Eq.~(\ref{eq:physical_matrix}) gives the vector
\begin{displaymath}
\bm{\eta}=\big[ b_{1}^i\ln \left( {1 - {}_A{\beta ^i}} \right), \ldots, b_{i - 1}^i\ln \left( {1 - {}_A{\beta ^i}} \right), b_{i+1}^i\ln \left( {1 - {}_A{\beta ^i}} \right), \ldots, b_{N}^i\ln \left( {1 - {}_A{\beta ^i}} \right)\big]^T,
\end{displaymath}
revealing the neighbors of node $i$ in the physical layer.

\section{Selection of threshold value for identification of existent links}

\begin{figure}[h]
\centering
\includegraphics[width=0.8\linewidth]{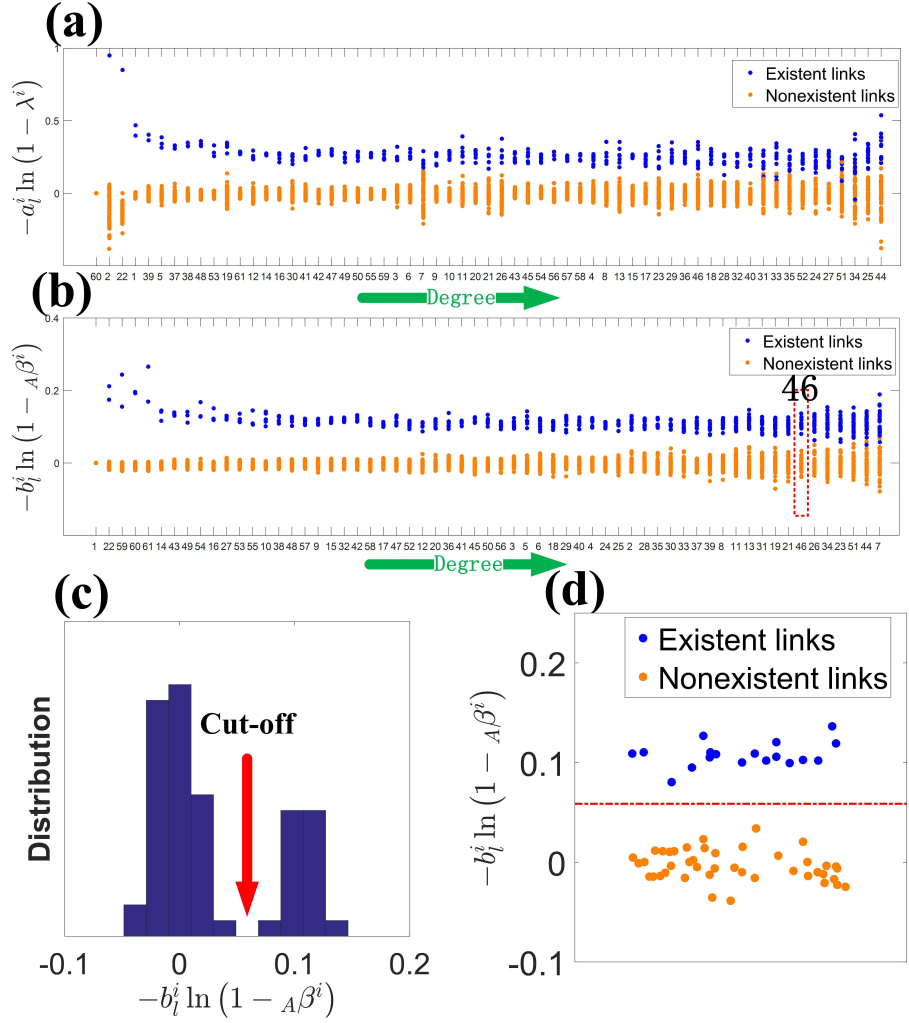}
\caption{ {\bf Reconstruction of CS-AARHUS duplex network}.
(a,b) Values of $-a_l^i\ln \left( {1 - {\lambda ^i}} \right), ~i\neq l$
and $-b_l^i\ln \left( {1 - {\lambda ^i}} \right)~i\neq l$ (b) for each node,
respectively. Each column gives the connectivity of a node. The blue and
orange points denote the existent and nonexistent links, respectively.
(c) Illustration of the choice of the threshold with node 46 [highlighted
by the red dashed frame in (b)]. Shown is the distribution of the values of
$-b_l^{46}\ln \left( {1 - {\lambda ^i}} \right)$ for $l\neq 46$. The peak
centered about zero corresponds to nonexistent links while the other peak
corresponds to existent links. A threshold can be set within the gap
between the two peaks. (d) The threshold is illustrated to distinguish the
actual from the nonexistent links. The length of time series is $M$=30000.
Other parameters are $\lambda=0.2$, ${}_U{\beta}=0.2$,
${}_A{\beta}=0.5{}_U{\beta}$, and $\mu=\delta=0.8$. }
\label{fig:cutoff}
\end{figure}

For each node $i$, the values of $a_l^i\ln \left( {1 - {\lambda ^i}} \right)$
(or of $b_l^i\ln \left( {1 - {}_A{\beta ^i}} \right)$) can be obtained from
Eq.~(\ref{eq:virtual_matrix}) [or Eq.~(\ref{eq:physical_matrix})]. From
Figs.~\ref{fig:cutoff}(a,b), we have that the values of
$-a_l^i\ln \left( {1 - {\lambda ^i}} \right)$
[or $-b_l^i\ln \left( {1 - {}_A{\beta ^i}} \right)$] are unequivocally above
zero for the actual links, while their values are close to zero for nonexistent
links, with a gap between the two sets of values. Representing the values
listed in each column as a histogram, we have that the peak centered about
zero corresponds to nonexistent links and the other corresponds to existent
links. A threshold value can be placed between the two peaks~\cite{SWFDL:2014},
as shown in Fig.~\ref{fig:cutoff}(c). A pair of nodes $i$ and $l$ are
connected if the corresponding value of
$-a_l^i\ln \left({1-{\lambda^i}}\right)$
[$-b_l^i\ln\left({1-{\lambda^i}}\right)$] is larger than the threshold.
Take node 46 as an example. We wish to infer its neighbors in the physical
layer [highlighted by the red dashed frame in Fig.~\ref{fig:cutoff}(b)].
Figure~\ref{fig:cutoff}(d) shows that the values larger than the threshold
correspond to the existent links.

\section{Reconstruction of duplex networks with heterogeneous rates of
spreading dynamics}

\begin{figure}[h]
\centering
\includegraphics[width=0.8\linewidth]{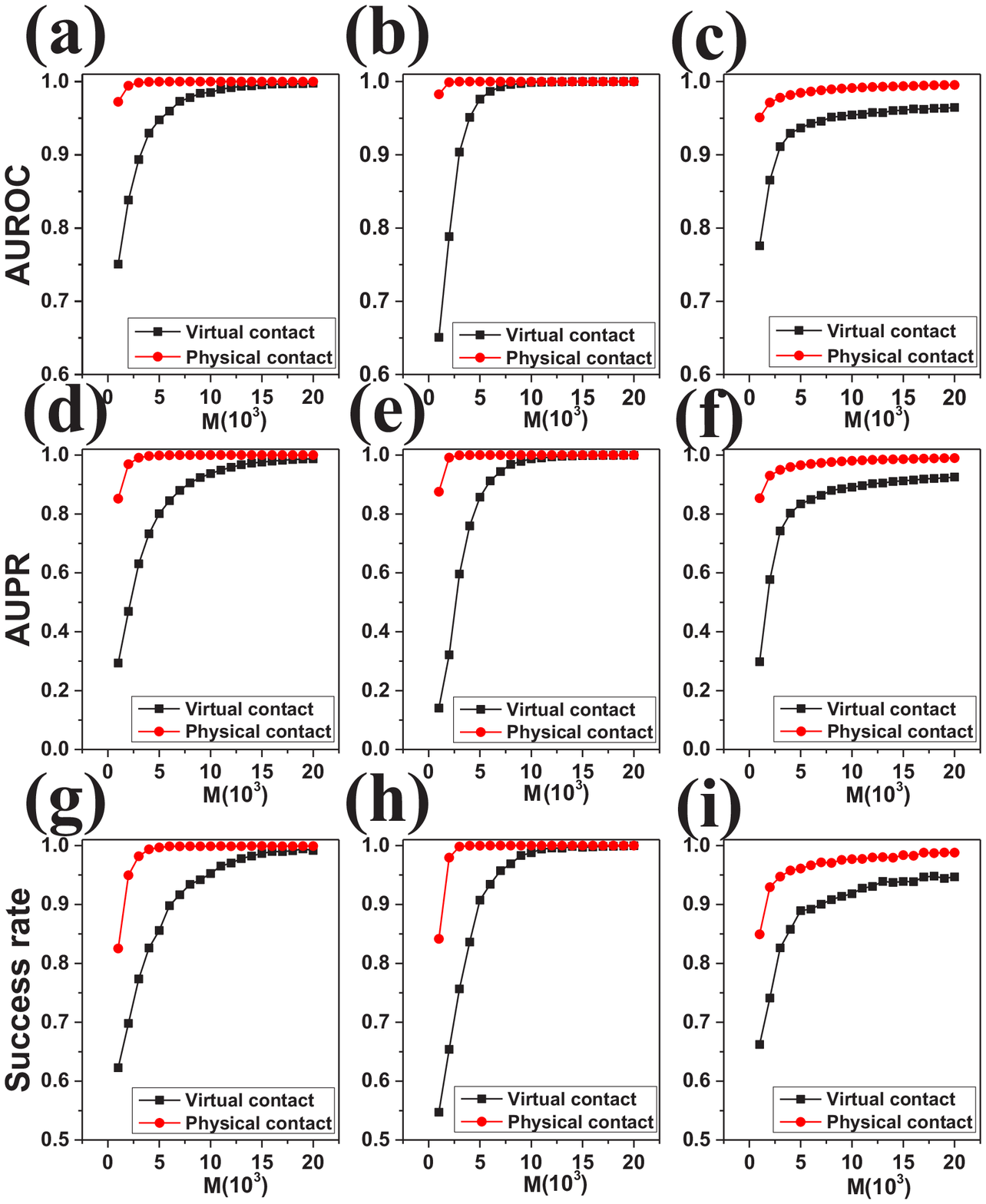}
\caption{ {\bf Applicability of reconstruction framework to spreading
dynamics with heterogeneous rates}. Reconstruction accuracy versus the
length $M$ of the time series for ER-ER (left column), SW-SW (central column)
and BA-BA (right column) duplex networks with heterogeneous transmission
and recovery rates. The network parameters are $N=100$ and
$\langle k_1\rangle=4,\langle k_2\rangle=6$.}
\label{fig:diveristy_rate}
\end{figure}

Figure~\ref{fig:diveristy_rate} demonstrates that our framework can
reconstruct duplex networks with heterogeneous rates of spreading dynamics.
In particular, transmission rates $\lambda^i$ and ${}_U{\beta ^i}$ are
randomly chosen from the ranges (0.2, 0.4) and (0.3, 0.5), respectively.
The recovery rates $\delta^i$ and $\mu^i$ are randomly picked up from the
ranges (0.6, 1) and (0.6, 1), respectively. Note that
${}_A\beta^i  = 0.5{}_U\beta^i$.

\end{document}